\documentclass[aps,prd,amsmath,amssymb,footinbib]{revtex4}
\usepackage{graphicx}
\usepackage{epsfig}
\usepackage{amsfonts}
\usepackage{amssymb}
\usepackage{epsf}
\newcommand{\insertplot}[5]{\begin{figure}
 \hfill\hbox to 0.05in{\vbox to #5in{\vfill
 \inputplot{#1}{#4}{#5}}\hfill}
 \hfill\vspace{-.1in}
 \caption{#2}\label{#3}
 \end{figure}}
\newcommand{\inputplot}[3]{
 \special{ps: plotfile #1}
}

\begin{document}

\title{On the Structure of Rotating Charged Boson Stars}

\author{Lucas G. Collodel}
 \email{lucas.gardai.collodel@uni-oldenburg.de}
\author{Burkhard Kleihaus}

 \author{Jutta Kunz}

\affiliation{
Institut f\"{u}r Physik, Universit\"{a}t Oldenburg, Postfach 2503 D-26111 Oldenburg, Germany}

\begin{abstract}
In this work we present full sets of solutions for rotating charged boson stars with different 
coupling values. By adopting local comoving coordinates, we are able to find expressions for 
the effective hydrodynamic quantities of the fluids as seen by this class of observers. 
We show that not only is the energy density non zero at the center, for the uncharged case 
it has a local maximum at the core from which it decreases until the point of local minimum 
where its variation is discontinuous.  For the first time, it is reported how rotating boson 
stars, charged and uncharged, are completely anisotropic fluids featuring three different 
pressures. Furthermore, the character of the electromagnetic fields is analyzed.
\end{abstract}

\maketitle

\section{Introduction}
Complex scalar fields minimally coupled to gravity, now coined boson stars, have been first 
realized half a century ago \cite{Kaup:1968zz,Feinblum:1968nwc}. One year later, it was shown 
that quantized real scalar fields give rise to the same equations of motion when treated 
semi-classically \cite{Ruffini:1969qy} and their gauged generalization, that endows the system 
with local $U(1)$ symmetry, came to be circa twenty years thereof \cite{Jetzer:1989av}. 
The more delicate case of rotating stars was found in the nineties 
\cite{Mielke:1997ag,Yoshida:1997qf} by solving the set of fully nonlinear partial differential 
equations after a perturbative approach for slow rotation \cite{Kobayashi:1994qi} turned fruitless. Thereafter, boson stars became subject of extensive studies of gravitational physics in the strong regime.

The initial setup of a free massive scalar field gave way to systems with interacting potentials 
that rendered more massive and astrophysically relevant stars, like the repulsive quartic 
interaction \cite{Colpi:1986ye} and the sixtic potential \cite{Friedberg:1986tq} for which 
solutions were named solitonic stars, for they would exist even in Minkowski spacetime as 
a self-binding soliton called Q-Ball \cite{Deppert:1979au,Coleman:1985ki}. The solitonic 
boson star has later been shown to give an appropriate description of what an axion 
star could be \cite{Barranco:2010ib}.

Rotating boson stars are extremely interesting for their unique properties which highly distinguish 
them from the more commonly investigated astrophysical objects such as black holes and neutron 
stars. For instance, their angular momentum is quantized \cite{Mielke:1997ag}, $J=\hbar mN$, 
where $m$ is a rotational integer and $N$ the particle number. Their full set of solutions, 
stability analysis, existence in higher dimensions and excited states are reported 
in \cite{Ryan:1996nk,Schunck:1996he,Schunck:1999pm,Mielke:2000mh,Kleihaus:2005me,Kleihaus:2007vk,Hartmann:2010pm,Kleihaus:2011sx,Collodel:2017biu}. Moreover, the topology of the scalar field changes upon rotation as it is then distributed along a torus as required by regularity. As a result, the maximum of their energy density happens off center warping spacetime in an unusual way, and the dynamics of particles freely falling in their spacetime takes a very peculiar form \cite{Grandclement:2014msa,Grould:2017rzz}. It has recently been shown \cite{Collodel:2017end} that when the $g_{tt}$ component of the metric contains a local maximum, which occurs in the surroundings of rotating boson stars, a ring of points is formed where particles initially at rest remain at rest due to an exclusive inertial phenomenon. 

In what concerns charged boson stars, their stability, quasi-bound states around black holes, 
behavior when critically charged and analytical approximations have broadly appeared in the 
literature \cite{Jetzer:1989us,Jetzer:1990wr,Jetzer:1993nk,Kleihaus:2009kr,Kleihaus:2010ep,Pugliese:2013gsa,Kan:2017rqk}. However, the general case when both rotation and charge are present has been less investigated. 
The system in absence of gravity is discussed in \cite{Radu:2008pp}, while some properties of rotating charged boson stars in an asymptotically flat spacetime described in \cite{Brihaye:2009dx}, and in four dimensional anti de-Sitter spacetime in \cite{Kichakova:2013sza}. The existence of solutions for hairy Kerr-Newman black holes has also been reported and analyzed in \cite{Delgado:2016jxq}.

In this paper we revisit the general case of charged rotating boson stars and construct full 
sets of solutions for different charges and rotational number. A new approach in terms of the 
hydrodynamic quantities of the fluids is given in local comoving coordinates, so that one can 
appreciate unambiguously how a certain class of observers measures the energy density and 
different pressures of the star and how they relate to each other. 

The paper is organized as follows. In Sec.~\ref{s1} we present the model that describes 
our system, obtain the partial differential equations and provide the necessary boundary 
conditions required for solving it and give the expressions for the conserved quantities 
that arise from the solutions. The numerical setup is briefly described in Sec.~\ref{s2}, 
where we give the results for the observables for the different parameter sets we solved for. 
We move to local coordinates in Sec.~\ref{s3} in order to define the energy density and 
pressures of our fluids, which are then calculated for a selected number of solutions. 
In Sec.~\ref{s4}, the invariants of the electromagnetic field are presented, 
in comparison to those of a Kerr-Newman black hole. For illustration purposes, 
we move to local zero angular momentum frames and calculate the components of the 
electric and magnetic fields. Our conclusions are drawn in Sec.~\ref{concl}.
The metric signature is taken to be $(-,+,+,+)$, 
Greek indices represent coordinate indices, while Latin indices
are used for the \emph{vierbein} basis. 
We use geometrical units such that $c=\hbar=8\pi G=1$.

\section{Theoretical Setting}
\label{s1}
In this section the action for charged boson stars and the corresponding
Einstein and field equations are presented. The ansatze for the metric, the boson field and 
the electro-magnetic potential are given and the boundary conditions for 
regular, asymptotically flat solutions are considered. Also, the conserved
charges and their interrelations are discussed.
\subsection{Action}
The system is described by a complex gauged scalar field minimally coupled to gravity,
\begin{equation}
\label{action}
S=\int \left\{\frac{R}{2}
-\frac{1}{2}g^{\mu\nu}\left[\left(D_\mu\Phi\right)\left(D_\nu\Phi\right)^{*}
                           +\left(D_\mu\Phi\right)^{\ast}\left(D_\nu\Phi\right)\right]
			   -U\left(\lvert\Phi\rvert\right)
			   -\frac{1}{4}F^{\mu\nu}F_{\mu\nu}\right\}\sqrt{-g}d^4x, 
\end{equation}
where $R$ is the curvature scalar, $\Phi$ is the complex scalar field, 
$U$ is the self-interaction potential of the scalar field, 
$F_{\mu\nu}=\partial_\mu A_\nu-\partial_\nu A_\mu$ is the electromagnetic field tensor and 
$D_\mu\equiv\nabla_\mu+i q A_\mu$ is the covariant derivative that minimally couples 
the scalar field to the gauge potential $A_\mu$ endowing the system with U(1) local symmetry.

Variation of  the action with respect to $g^{\mu\nu}$ leads to the 
Einstein's field equations, 
\begin{equation}
\label{equEinst}
G_{\mu\nu}\equiv R_{\mu\nu}-\frac{1}{2}Rg_{\mu\nu}=T_{\mu\nu} \ ,
\end{equation}
where the energy momentum tensor 
reads
\begin{align}
T_{\mu\nu} = &
-g_{\mu\nu}\left\{\frac{1}{2}g^{\sigma\lambda}
\left[\left(D_\sigma\Phi\right)\left(D_\lambda\Phi\right)^{*}
     +\left(D_\sigma\Phi\right)^{\ast}\left(D_\lambda\Phi\right)\right]
     +U\left(\lvert\Phi\rvert\right)
     +\frac{1}{4}F^{\sigma\lambda}F_{\sigma\lambda}\right\}
     \\ \nonumber
 &     
+\left[\left(D_\mu\Phi\right)^{\ast}\left(D_\nu\Phi\right)
       +\left(D_\mu\Phi\right)\left(D_\nu\Phi\right)^{\ast}\right]
       +F_{\mu\sigma}F_{\nu\lambda}g^{\sigma\lambda}.
\end{align}
The field equations of the scalar field are obtained by variation of the Lagrangian
with respect to $\Phi^{*}$
\begin{equation}
\label{equPhi}
D^\mu D_\mu \Phi = \frac{\partial U}{\partial |\Phi|^2} \Phi \ , 
\end{equation}
and variation of the Lagrangian with respect to the gauge potential yields the Maxwell
equations
\begin{equation}
\label{equMax}
\nabla_\nu F^{\nu\mu} = q j^{\mu}
\end{equation}
with conserved electro-magnetic current
\begin{equation}
\label{equJelm}
j^{\mu}=-i(\Phi^{\ast}\partial^{\mu}\Phi-\Phi\partial^{\mu}\Phi^{\ast})+2q|\Phi|^2A^\mu, 
\qquad 
\nabla_\mu j^{\mu} = 0 \ .
\end{equation}
\subsection{Ansatze} 
We are interested in stationary axisymmetric boson star solutions. We adopt the quasi-isotropic 
Lewis-Papapetrou metric in adapted spherical coordinates $(t,r,\theta,\varphi)$, for 
which the line element reads
\begin{equation}
\label{metric}
ds^2=-fdt^2
+\frac{l}{f}\left[g\left(dr^2+r^2d\theta^2\right)
                 +r^2\sin^2\theta\left(d\varphi-\frac{\omega}{r}dt\right)^2\right] \ ,
\end{equation}
where metric functions $f$, $l$, $g$ and $\omega$ are functions of $r$ and $\theta$ only. 
This spacetime then possesses two Killing vector fields, 
namely $\xi=\partial_t$, and $\eta=\partial_\varphi$. 
In this spacetime, the only off diagonal non-zero components of Einstein's tensor are 
$G_{t\varphi}$ and $G_{r\theta}$ (and their symmetric counter parts), from which it follows 
that the gauge potential has only two non-trivial contributions, 
$A=V(r,\theta)dt+C(r,\theta)d\varphi$. 
The usual Ansatz for the boson field \cite{Mielke:1997ag}
\begin{equation}
\label{ansatz}
\Phi(t,r,\theta,\varphi)=\phi(r,\theta)e^{i\omega_st+i m\varphi},
\end{equation}
fixes the gauge. 
Here the boson frequency $\omega_s$ and the winding number $m$ are real constants and furthermore 
$m$ is an integer due to the identification $\Phi(\varphi)=\Phi(\varphi+2\pi)$.
One could instead work with a real scalar field, and the parameters $\omega_s$ and $m$ 
would then appear in the boundary conditions for $A_\mu$. On the other hand, 
there is no single valued function $h(x^\mu)=\int A_\mu dx^\mu$ for which the gauge 
transformation $\Phi\rightarrow\Phi e^{ih(x^\mu)}$, 
$A_\mu\rightarrow A_\mu-\frac{i}{q}\partial_\mu h(x^\mu)$ makes $A_\mu$ trivial everywhere 
for $F_{\mu\nu}\neq 0$. It is worth noticing that there is a screening mechanism in this system 
due to the $\phi^2 A_\mu A^\mu$ term in the Lagrangian, which corresponds to a position dependent mass term.

In the absence of gravity, this is the simplest interacting gauge theory one can write down. 
Indeed, if one is interested in Q-balls, 
which are bound through their self-interaction, one must adopt non-renormalizable 
potentials \cite{Deppert:1979au,Mielke:1980sa,Volkov:2002aj}. 
It was pointed out in \cite{Barranco:2010ib}, based on the pioneering work \cite{Ruffini:1969qy}, 
that real quantized scalar fields yield the same equations of motion as those of complex 
scalar fields in an entirely classical approach. In this sense, the axion which is described by a 
real quantized scalar field would not give rise to the oscillatons \cite{UrenaLopez:2001tw} when 
coupled to gravity, which are time dependent solutions of a gravitationally bound real scalar 
field obtained through the classical approach. Instead, when treated classically, axion stars are 
realized by solitonic boson stars, e.g. a complex scalar field minimally coupled to gravity. 
The axion potential is given by
\begin{equation}
\label{axionpotential}
U_a(\Phi)=m_af_a\left[1-\cos\left(\frac{\Phi}{f_a}\right)\right],
\end{equation}
where $m_a$ is the axion mass and $f_a$ is the decay constant. In the semi-classical approach, 
the field is quantized, $\Phi\rightarrow\hat{\Phi}=\hat{\Phi}^++\hat{\Phi}^-$, and in order to 
appreciate the action of the potential on the different states and calculate the expectation 
value of energy-momentum tensor $\langle T^\mu{}_\nu\rangle$, the self-interaction potential is 
Taylor expanded,
\begin{equation}
\label{tayloraxionpotential}
U_a(\Phi)\sim
\frac{m_a^2}{2}\Phi^2-\frac{1}{4!}\frac{m_a^2}{f_a^2}\Phi^4+\frac{1}{6!}\frac{m_a^2}{f_a^4}\Phi^6-....
\end{equation}
It is shown in \cite{Barranco:2010ib} that the solutions do not depend strongly on the number 
of terms considered in the above expansion, as long as the quartic term is present with the 
correct minus sign. Because we are interested in a theory with a lower bound for the energy,
we consider yet the next term in the expansion above adopting the sixtic potential, which also gives rise
to Q-balls in the absence of gravity,

\begin{equation}
U\left(\lvert\Phi\rvert\right)=\phi^2\left(m_b^2-a\phi^2+b\phi^4\right),
\end{equation}
with $m_b^2=1.1$, $a=2$ and $b=1$. We will keep the boson frequency $\omega_s$, the
gauge coupling parameter $q$ and the winding number $m$ as free parameters, which 
determine the charged rotating boson star solutions.
\subsection{Boundary Conditions}
When the ansatze for the metric, the scalar field and the gauge field are substituted in 
Eqs.~(\ref{equEinst}), (\ref{equPhi}) and (\ref{equMax}) the general field equations
reduce to a system of coupled nonlinear partial differential equations in $r$ and $\theta$.
In order to find asymptotically flat and regular solutions boundary conditions
need to be imposed on the functions, respectively their normal derivatives,
at the origin and in the asymptotic region, as well as along the symmetry axis.
Also, for solutions with even parity boundary conditions in the equatorial plane are imposed.

At the origin, regularity requires that
\begin{align}
\partial_rf\rvert_{r=0}=0,\qquad 
\partial_rl\rvert_{r=0}=0,\qquad 
g\rvert_{r=0}=1,\qquad 
\omega\rvert_{r=0}=0, \qquad 
\phi\rvert_{r=0}=0, \qquad 
\partial_rV\rvert_{r=0}=0, \qquad 
C\rvert_{r=0}=0.
\end{align}
Since our spacetime is asymptotically Minkowski, we need the scalar and electromagnetic fields 
to be zero at spatial infinity,
\begin{align}
f\rvert_{r\rightarrow\infty}=1,\qquad 
l\rvert_{r\rightarrow\infty}=1,\qquad 
g\rvert_{r\rightarrow\infty}=1,\qquad 
\omega\rvert_{r\rightarrow\infty}=0, 
\qquad \phi\rvert_{r\rightarrow\infty}=0, 
\qquad V\rvert_{r\rightarrow\infty}=0, 
\qquad C\rvert_{r\rightarrow\infty}=0,
\end{align}
where we stress again that the values of $V$ and $C$ are set by the gauge.

On the symmetry axis, the elementary flatness conditions sets $g\rvert_{\theta=0}=1$. 
The other fields are, once again, determined as to guarantee regularity,
\begin{align}
\partial_\theta f\rvert_{\theta=0}=0,\qquad 
\partial_\theta l\rvert_{\theta=0}=0,\qquad 
g\rvert_{\theta=0}=1,\qquad 
\partial_\theta\omega\rvert_{\theta=0}=0, 
\qquad \phi\rvert_{\theta=0}=0, 
\qquad \partial_\theta V\rvert_{\theta=0}=0, 
\qquad C\rvert_{\theta=0}=0,
\end{align}
and one can appreciate how rotation brings the non-trivial scalar field 
to possess non-trivial topology. 

Finally, because we are describing a system with even parity, all angular derivatives must 
vanish on the equatorial plane,
\begin{align}
&\partial_\theta f\rvert_{\theta=\pi/2}=0,\qquad 
\partial_\theta l\rvert_{\theta=\pi/2}=0,\qquad 
\partial_\theta g\rvert_{\theta=\pi/2}=0,\qquad 
\partial_\theta\omega\rvert_{\theta=\pi/2}=0,
\\ \nonumber
&\partial_\theta\phi\rvert_{\theta=\pi/2}=0, \qquad 
\partial_\theta V\rvert_{\theta=\pi/2}=0, \qquad 
\partial_\theta C\rvert_{\theta=\pi/2}=0.
\end{align}
\subsection{Conserved Quantities}
Charged boson stars are characterized by physical observables like mass $M$, angular 
momentum $J$, electric charge $Q$, respectively particle number $N$, 
as well as dipole moment $\mu$.
Here we discuss the interrelation of these quantities and how they are obtained.
In a stationary asymptotically flat spacetime, the Komar expression provides a way to calculate 
global quantities directly associated with the Killing vectors. The mass and the angular momentum
\begin{equation}
M=2\int_\Sigma R_{\mu\nu}n^\mu\xi^\nu dV,\qquad J=-\int_\Sigma R_{\mu\nu}n^\mu\eta^\nu dV,
\end{equation} 
are then calculated as an integral over a spacelike asymptotically 
flat hypersurface  $\Sigma$ bounded at spatial infinity. 
Here, $R_{\mu\nu}$ is the Ricci tensor,  
$n^\mu$ is a vector normal to $\Sigma$ with $n^\mu n_\mu=-1$, 
and $dV=\sqrt{-g/f}drd\theta d\varphi$ denotes the natural volume element.
The metric (\ref{metric}) implies that $n^{\mu}=(\xi^\mu+\omega/r\eta^\mu)/\sqrt{f}$. 
Expressing the Ricci tensor in terms of the energy-momentum tensor and employing the field 
equations, yields 
\begin{equation}
\label{intmj}
M=\int(2T^t_{\hphantom{a}t}-T)\sqrt{-g}drd\theta d\varphi, \qquad 
J=-\int T^t_{\hphantom{a}\varphi}\sqrt{-g}drd\theta d\varphi \ .
\end{equation}

Local gauge symmetry gives rise to a conserved Noether current $j^{\mu}$ (Eq.~(\ref{equJelm}).
The associated electric charge $Q$ can be obtained by integrating the projection of the 
current onto the future directed hypersurface normal $n^\mu$ and integrating over the 
whole space,
\begin{align}
\label{intq}
N &= \int_\Sigma  j_\mu n^\mu dV, \qquad Q=qN,  
\end{align}
where $N$ is considered as particle number of the star.

The integrand of the above equation is simply $-j^t\sqrt{-g}$. It is well known that for 
uncharged rotating boson stars $T^t_{\hphantom{a}\varphi}=mj^t$ and the angular momentum is 
quantized, assuming always values that are multiples of the particle number 
$J=mN$ \cite{Mielke:1997ag}. The charged case is more involved. 
Here we show that the quantization relation valid for uncharged boson stars 
also holds for charged boson stars.
We note that the integrand of the angular 
momentum can be written as
\begin{equation}
\label{ttp}
T^t{}_\varphi=m j^t +qA_\varphi j^t+F^{t\mu}F_{\varphi\mu} \ .
\end{equation}
The third term can be rewritten as
\begin{equation}
F^{t\mu}F_{\varphi\mu}=F^{t\mu}\nabla_\varphi A_\mu-\nabla_\mu\left(F^{t\mu}A_\varphi\right)-qA_\varphi j^t,
\end{equation}
where the Maxwell equation (\ref{equMax}) has been used.
The first term on the right hand side is identically zero, the second one is a total 
divergence, which does not contribute to the integral ($A_\varphi=0$ at infinity) 
and the third one cancels the second term  in Eq.~(\ref{ttp}).
Even though the integrands of $J$ and $N$ are 
different, after integration the quantisation relation $J=mN$ still holds in the 
rotating charged case.

The global quantities $M$, $J$, $Q$, and furthermore the magnetic moment $\mu$ can be extracted 
from the asymptotic behaviour of the metric and gauge field functions, as 
\begin{equation}
\label{asympmj}
M=\frac{1}{2}\lim_{r\rightarrow\infty}r^2\partial_r f, \qquad 
J=\frac{1}{2}\lim_{r\rightarrow\infty}r^2\omega, \qquad 
Q=\frac{1}{2}\lim_{r\rightarrow\infty}r^2\partial_r V, \qquad 
\mu=\frac{1}{2}\lim_{r\rightarrow\infty}r^2\partial_r C.
\end{equation} 
\section{Numerical Solutions}
\label{s2}
The system comprises seven coupled nonlinear partial differential equations to be solved for four 
metric functions ($f,l,g$ and $\omega$) and three matter/gauge functions ($\phi, V$ and $C$). 
In order to solve this system numerically, we employ a two dimensional grid with a compactified 
radial coordinate $x=r/(r+1)$ where $x\in[0,1]$ covers the radial direction from zero to infinity, 
and $\theta\in[0,\pi/2]$ since all of the quantities have even parity with respect to reflections 
at the equatorial plane. The equations are then solved with the subroutines for elliptical 
PDEs of the FIDISOL package, with most grids containing $125\times 50$ points and precision 
of $10^{-7}$.
\subsection{Domain of Existence} 
The charged rotating boson stars are determined by the 
boson frequency $\omega_s$, the gauge coupling parameter $q$ and the winding number $m$.
We keep the winding number fixed,  $m=1$. For $q=0$ the uncharged rotating boson stars are obtained.
The charged rotating boson stars emerge from
the uncharged rotating boson stars when the parameter $q$ is increased.
In the non-rotating case the range of the parameter $q$ was discussed in \cite{Jetzer:1989us,Pugliese:2013gsa}.
In our study of rotating boson stars we found that solutions exist up to
the maximal value of $q$. However, numerics fails when the maximal value is approached. 

As for the uncharged rotating (and non-rotating)
boson stars the range of $\omega_s$ is restricted to  
$\omega_{s, \,{\rm min}} < \omega_s < \omega_{s, \,{\rm max}} $,
where $\omega_{s, \,{\rm max}}= m_b$. The minimal value $\omega_{s, \,{\rm min}}$ has to be 
determined numerically and depends on the gauge coupling parameter $q$. 
We observe that $\omega_{s, \,{\rm min}}$ increases with increasing values of $q$,
leading to a restricted range of the frequency  $\omega_s$.
\subsection{Observables} 
For convenience we will use the quantity $\phi_1 =\partial_r \phi\rvert_{r=0}$ instead 
of $\omega_s$ as parameter. $\phi_1=0$ corresponds to $\omega_s = \omega_{s, \,{\rm max}}$.

\begin{figure}[h!]
\begin{center}
\mbox{\includegraphics[width=0.4\textwidth, angle =0]{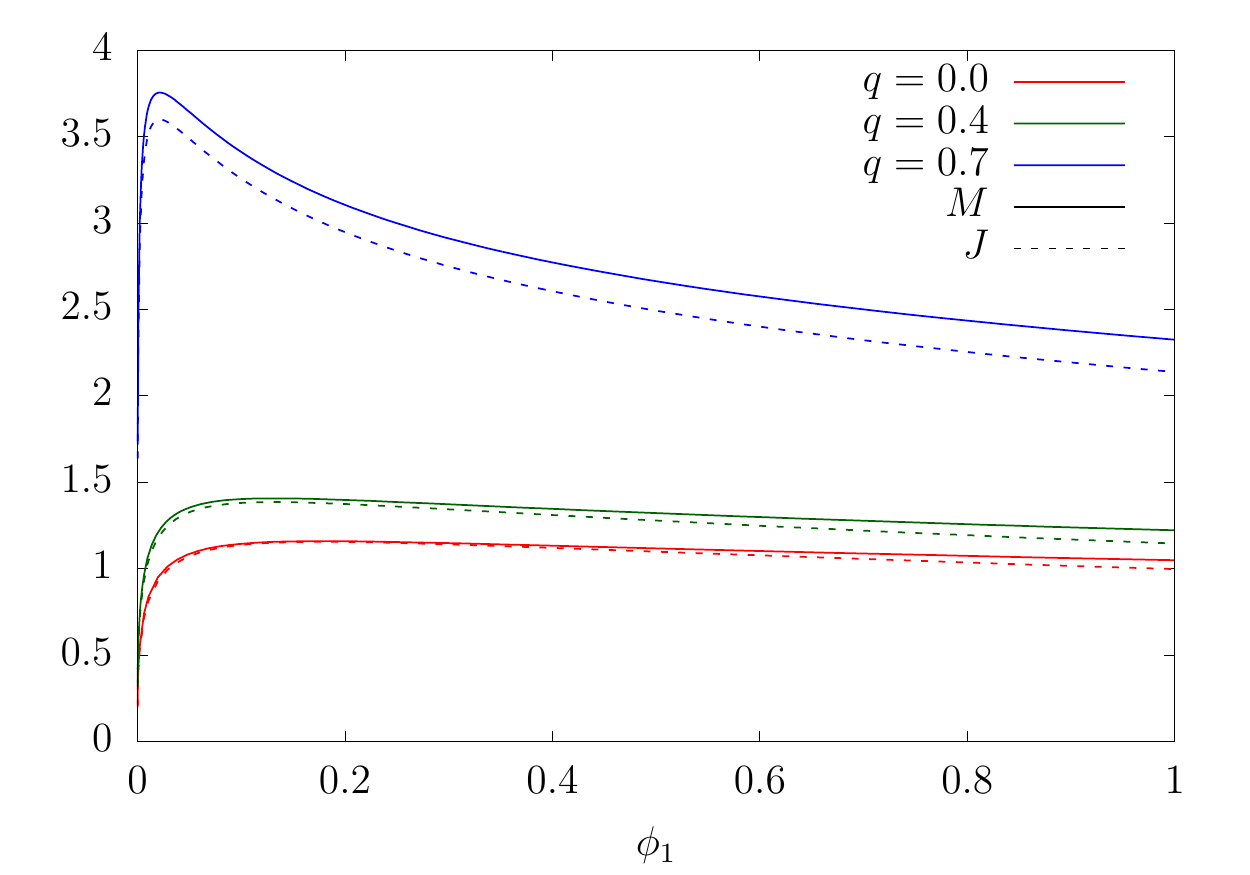}}
\mbox{\includegraphics[width=0.4\textwidth, angle =0]{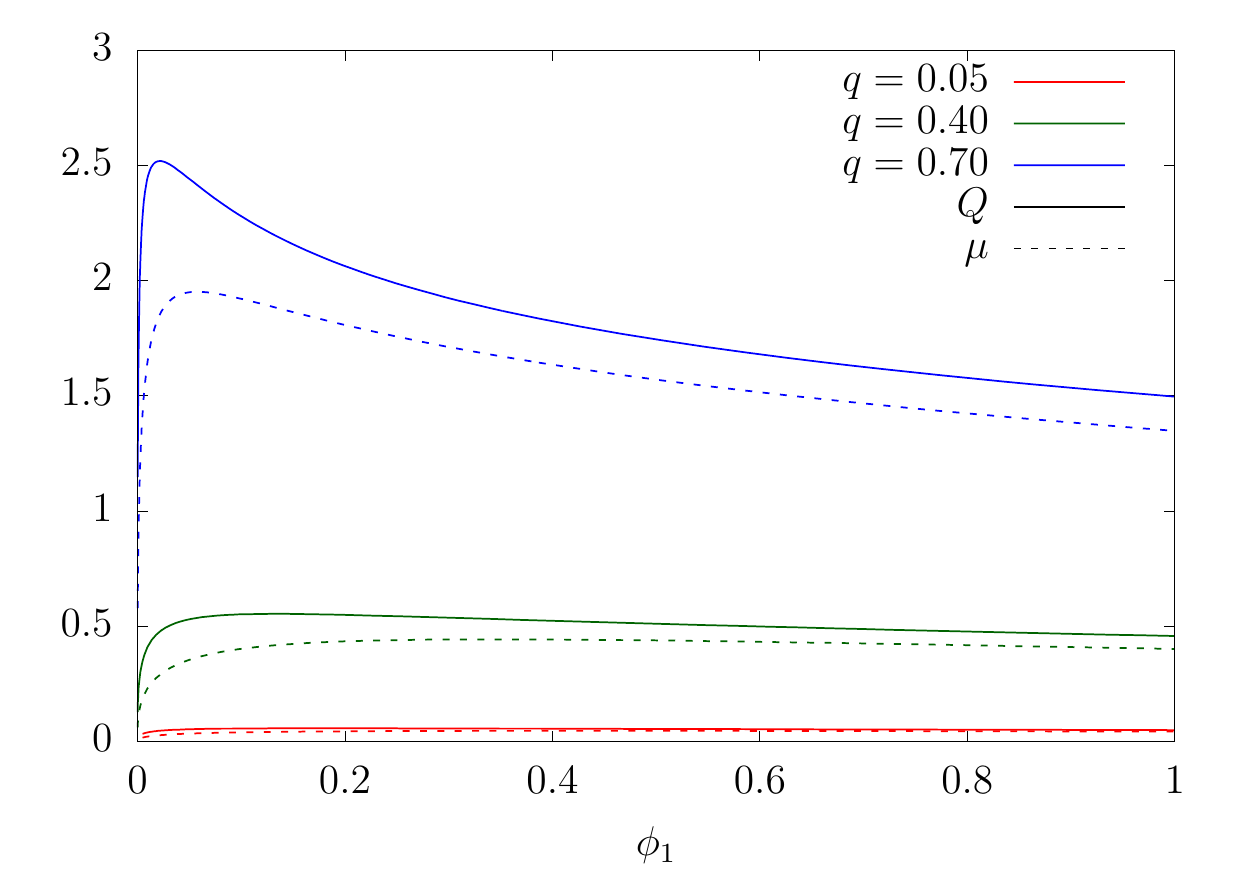}}
\end{center}
\caption{\emph{Left:} Mass and angular momentum for rotating stars with $m=1$, 
uncharged $q=0$ and charged with $q=0.4$ and $q=0.7$. 
\emph{Right:} Charge and magnetic moment for three families of solutions with $m=1$: 
$q=0.05$, $q=0.40$, $q=0.70$. }
\label{mjqu}
\end{figure}

The mass, angular momentum, charge and magnetic moment as function of $\phi_1$ are 
presented in Fig. \ref{mjqu} for some family of solutions. We observe  that the mass and angular 
momentum increase with increasing charge for fixed value of $\phi_1$.
The maximum mass and angular momentum solution occurs for smaller values of $\phi_1$ as 
the charge increases.  As in the non-rotating case, the Coulomb repulsion between the 
star's components accounts to an equilibrium state at a higher mass value for the same value 
of $\phi_1$ when compared to solutions with smaller gauge couplings.


%

\begin{figure}[h!]
\begin{center}
\mbox{\includegraphics[width=0.4\textwidth, angle =0]{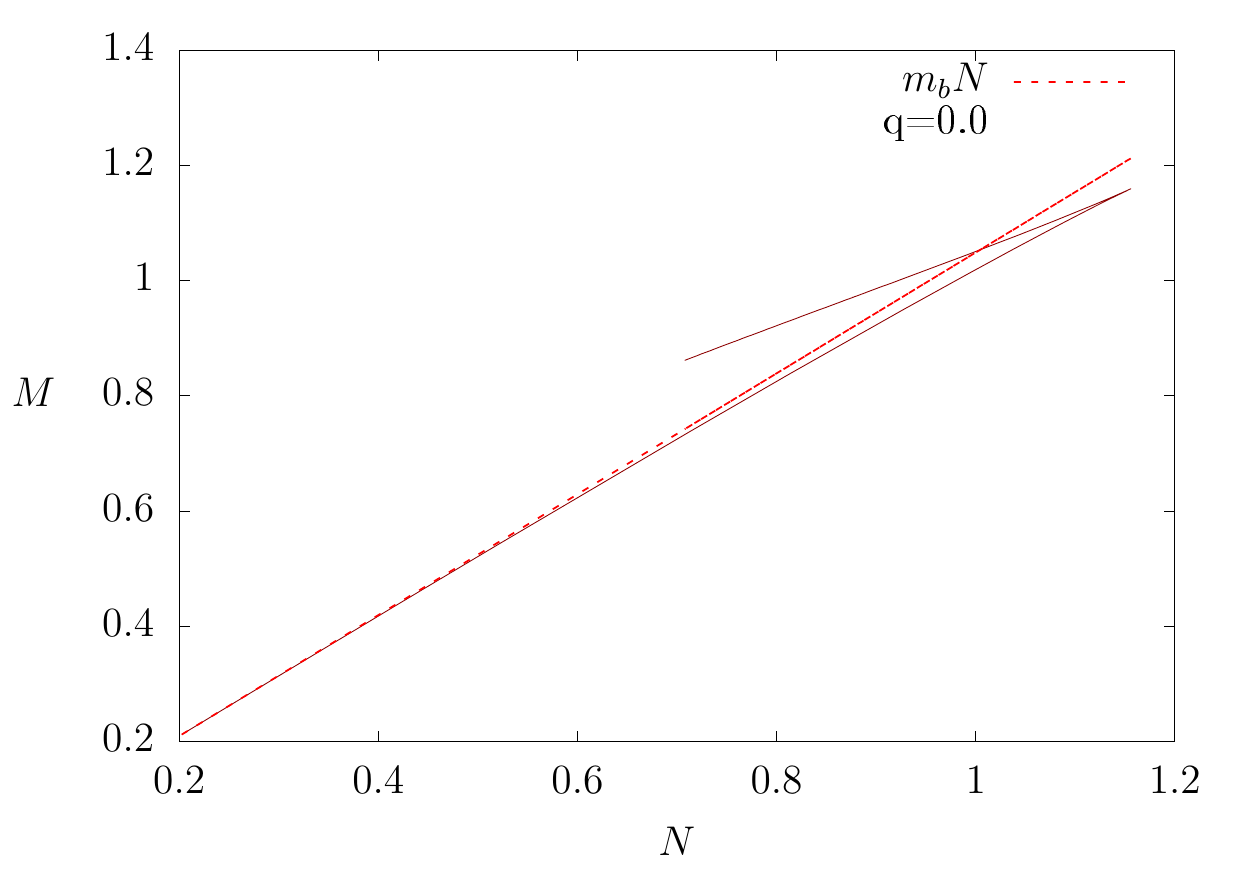}}
\mbox{\includegraphics[width=0.4\textwidth, angle =0]{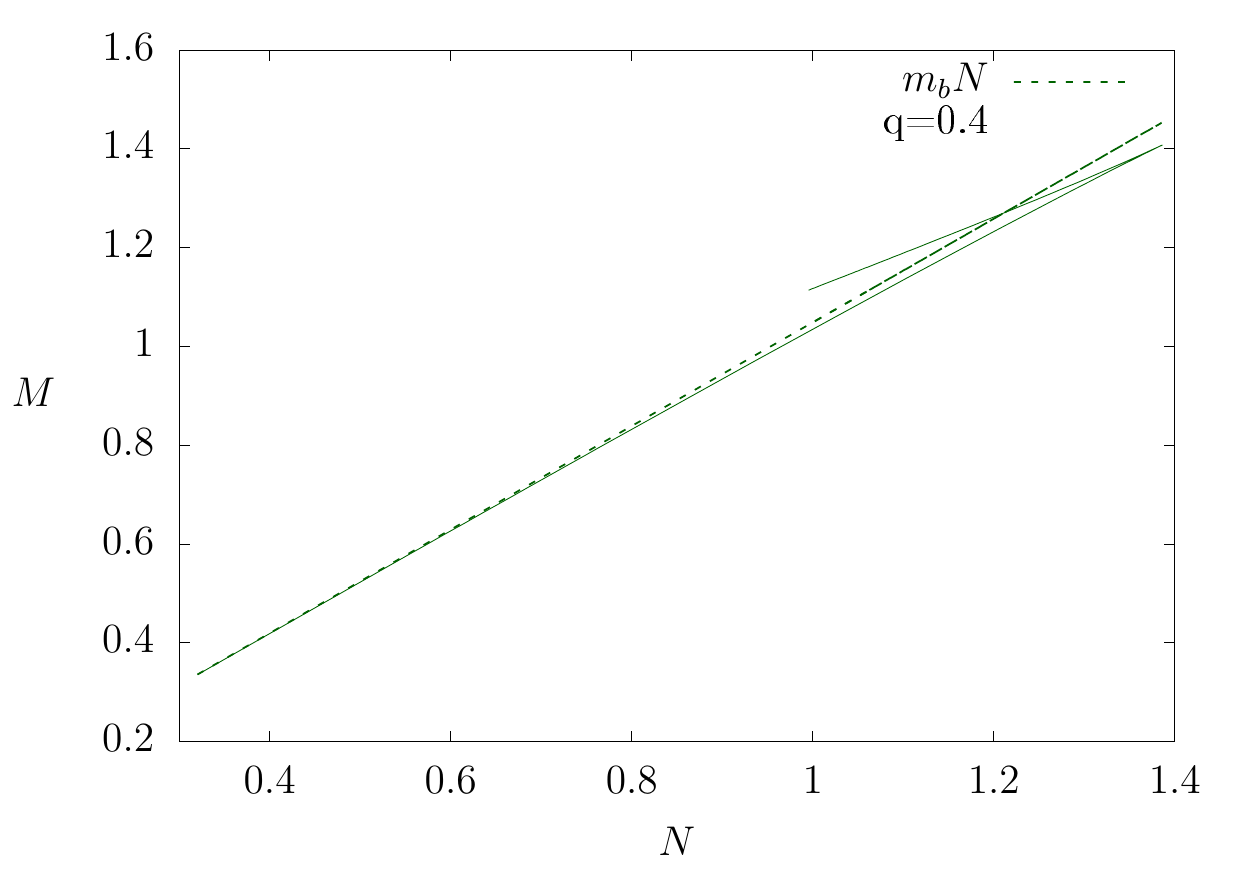}}
\mbox{\includegraphics[width=0.4\textwidth, angle =0]{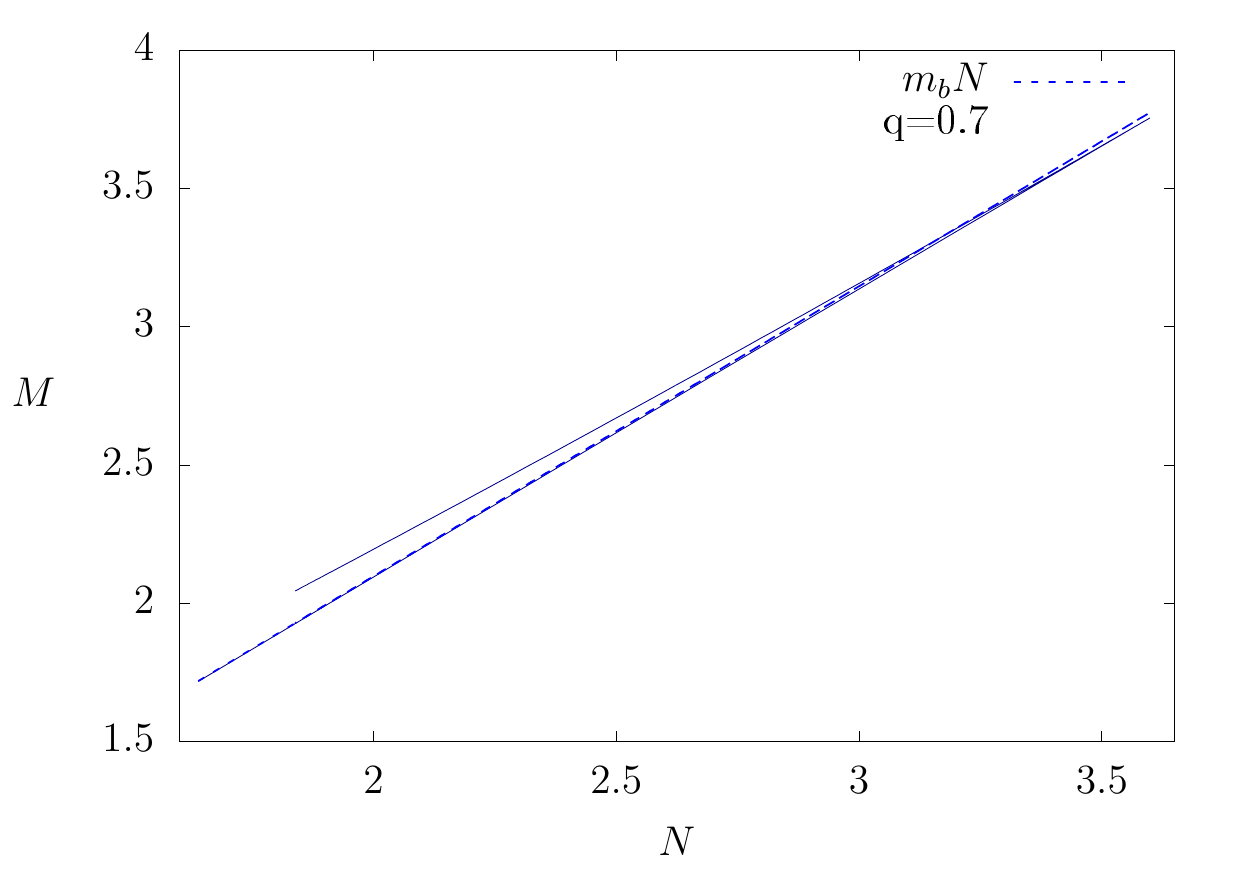}}
\end{center}
\caption{Mass vs. particle number for $m=1$ and gauge coupling parameter
$q=0.0$, $q=0.4$ and $q=0.7$.
}
\label{qxm}
\end{figure}

In order to have some insight in the stability of these solutions, 
we refer to the diagrams in Fig. \ref{qxm}, where we show the mass as function
of the particle number for fixed values of the gauge coupling parameter, together
with the line representing $N$ free bosons with mass $m_b$.
The qualitative behavior is the same as described in \cite{Kleihaus:2011sx,Collodel:2017biu}. 
The first branch comprehends the solutions with non-topological stability. It extends up to
the maximal mass and maximal particle number, where it connects to a second branch which
extends back to smaller masses and particle number. Along the second branch the solutions
possess larger mass at given particle number than the solutions on the first branch. 
Thus the solutions on the second branch are quantum mechanically unstable.
Once the mass on the second branch exceeds the mass of free bosons (at given particle number)
the boson stars are also classically unstable.

The new feature we observe is that
the higher the value of the gauge coupling, the smaller is the difference between 
the mass of the boson star and the mass of free bosons with the same particle number.

The larger the  gauge coupling, the larger the Coloumb repulsion of the
particles of the star and the more its mass resembles that of a gas of separated particles.


%

\subsection{Static Ring}

In a previous paper \cite{Collodel:2017end}, we have shown that a class of spacetimes contains 
what we called the \emph{static ring}: a ring of points in the equatorial plane, 
centered at the origin, where a particle initially at rest remains at rest. 
This class of spacetimes is assumed to be stationary, axisymmetric and circular. 
A necessary and sufficient condition for the existence of such ring is that the $g_{tt}$ 
component of the metric possesses a local maximum at a point where it is negative, 
i.e. not in an ergoregion. Boson stars and highly compact objects surrounded by a 
massive bosonic cloud are the most prominent sources of this class of spacetimes and 
the rotating charged boson star is no different. 

\begin{figure}[h!]
\begin{center}
\mbox{\includegraphics[width=0.4\textwidth, angle =0]{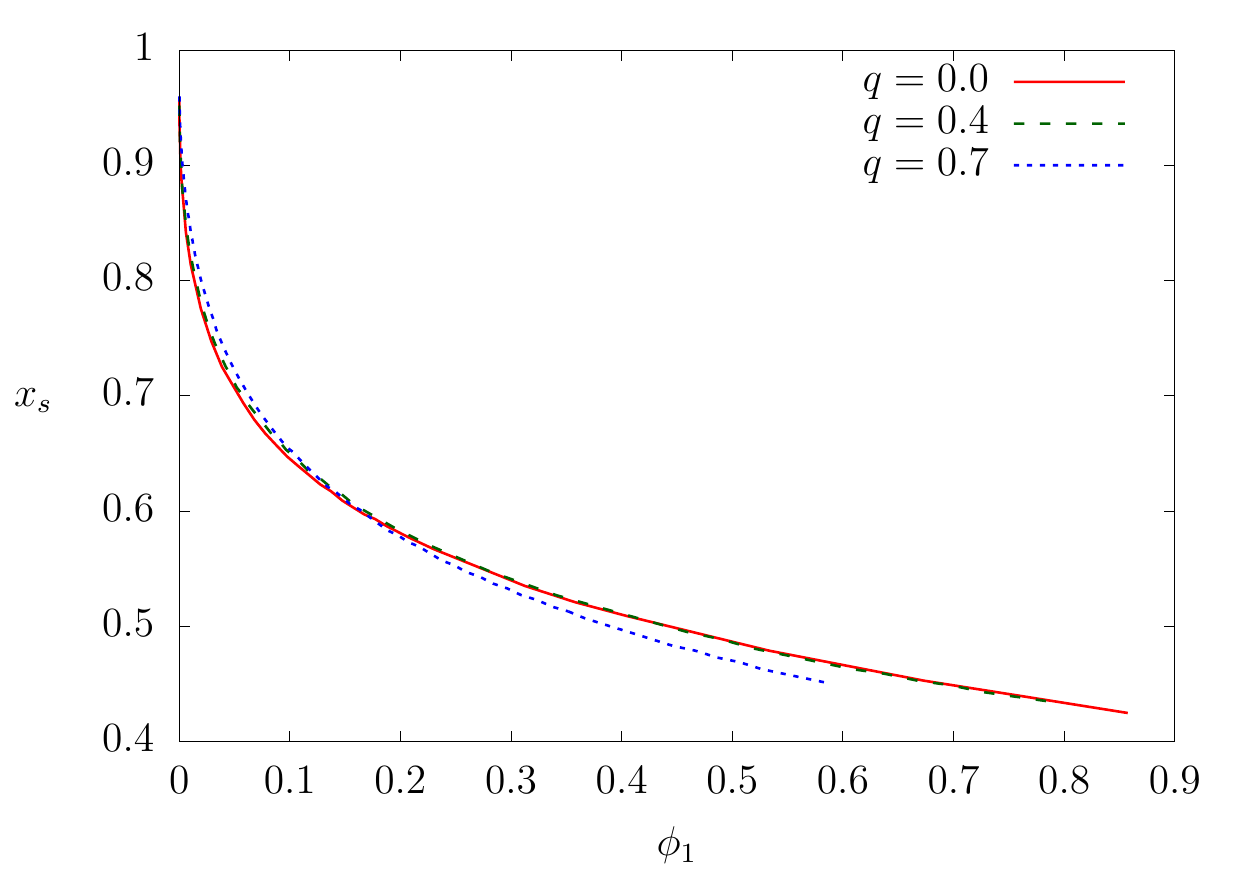}}
\end{center}
\caption{
Location of the static ring: the compactified coordinate $x_s$ of the static ring is shown as 
function of the parameter $\phi_1$. 
}
\label{staticring}
\end{figure}

The radius of the static ring for three different family of solutions is illustrated in 
Fig. \ref{staticring}. All of the solutions, for every values of the gauge coupling 
parameter and $\phi_1$, the function $g_{tt}$ shows qualitatively the same behaviour, 
containing a local maximum in the equatorial plane. 
Fig. \ref{staticring} shows the compactified coordinate of the static rings as
function of the parameter  $\phi_1$ for fixed gauge coupling parameter $q$.
The curves terminate at a specific point where the static ring enters the
ergoregion, in which no timelike particle could stay at rest. 

As the gauge coupling increases, the radius of static ring decreases slightly and the 
curves  terminate at smaller values of $\phi_1$.
Hence the radius of the static ring does not depend strongly on the 
gauge coupling parameter. 

The more distinctive behavior change as the coupling increases is the terminating point,
which  happens for smaller values of $\phi_1$. Since the angular momentum $J$ increases 
due to a less latent gauge field, $A_\mu$, ergoregions are prone to appear earlier in the
parameter space, even though all sets are described by the same rotation number.


%

\section{Comoving Observer}
\label{s3}
In order to have a clear realization of the energy density and pressures separately, 
we adopt comoving coordinates, i.e. for which the energy momentum tensor is diagonal. 
These quantities are then the eigenvalues that satisfy
\begin{equation}
T^\mu_{\phantom{a}\nu}\hat{e}^\nu_{\phantom{a}(a)}=\lambda_{(a)}\hat{e}^\mu_{\phantom{a}(a)},
\end{equation}
where $\hat{e}^\mu_{\phantom{a}a}$ are the energy momentum tensor's eigenvectors. 

The eigenvalues and eigenvectors found through the equation above are rather long for the 
system we are describing. Nevertheless, for any circular, stationary and axisymmetric spacetime, 
we can hide the extensive expressions in notation by splitting the energy momentum tensor as
\begin{equation}
T_{\mu\nu}=\mathcal{T}_{\mu\nu}^{(1)}+\mathcal{T}_{\mu\nu}^{(2)},
\end{equation}
with
\begin{equation}
  \mathcal{T}_{\mu\nu}^{(1)} = 
 \left(
 \begin{array}{cccc}
 T_{tt}        & 0 &  0 &  T_{t\varphi}\\
 0             & 0 &  0 & 0 \\
 0             & 0 &  0 & 0 \\
 T_{t\varphi}  & 0 &  0 & T_{\varphi\varphi}
\end{array}
\right) \ , \ \ \ 
  \mathcal{T}_{\mu\nu}^{(2)} = 
 \left(
 \begin{array}{cccc}
 0     & 0           &  0                & 0\\
 0     & T_{rr}      &  T_{r\theta}      & 0 \\
 0     & T_{r\theta} &  T_{\theta\theta} & 0 \\
 0     & 0           &  0                & 0 
\end{array}
\right)\nonumber
\end{equation}
Then the eigenvalues of $T_{\mu\nu}$ are given by
\begin{equation}
\label{eigenvalues}
\lambda_{i\pm}=\frac{1}{2}\left[\pm\sqrt{2\mathcal{T}^{\mu\nu}\mathcal{T}_{\mu\nu}
                                -\mathcal{T}^2}+\mathcal{T}\right]_{(i)},
\end{equation}
where $\mathcal{T}^{(i)} =g^{\mu\nu} \mathcal{T}_{\mu\nu}^{(i)}$.
\subsection{Non-rotating Boson Stars}
In the interest of comparison, we shall consider the
spherically symmetric charged boson stars, again employing isotropic coordinates.
In this case $\omega=0$, $g=1$ and $C=0$ and $f$, $l$, $\phi$ and $V$ are functions of
$r$ only. The energy density and pressures then read
\begin{align}
\rho   &=\frac{f}{l}\phi'^2
        +\frac{V'^2}{2l}+\frac{\phi^2}{f}\left(Vq+\omega_s\right)^2+U\left(|\Phi|\right), 
\nonumber \\
p_r    &=\rho-\frac{V'^2}{l}-2U\left(|\Phi|\right), 
\nonumber \\
p_\bot &=\rho-2\frac{f}{l}\phi'^2-2U\left(|\Phi|\right).
\end{align}
The uncharged case is obtained simply by $q=0$ and $V=0$.
\subsection{Rotating Boson Stars}

The rotating case is of course more involved. Let us first consider the uncharged 
rotating boson stars. The hydrodynamic quantities obtained 
via eq. (\ref{eigenvalues}) are
\footnote{In \cite{Collodel:2017biu} we have overlooked that a term should actually be 
an absolute value and reported incorrect expressions.}.
\begin{align}
\label{rbshyd}
\rho &=\frac{f}{lgr^2}
\left[
       r^2\left(\partial_r\phi\right)^2+\left(\partial_\theta\phi\right)^2
\right]
      +\frac{\left|f^2m^2-l\sin^2\theta(m\omega+\omega_sr)^2\right|}{flr^2\sin^2\theta}\phi^2
      +U\left(|\Phi|\right), 
\nonumber \\
p_r &=\frac{f}{lgr^2}
\left[
       r^2\left(\partial_r\phi\right)^2+\left(\partial_\theta\phi\right)^2
\right]
      -\frac{f^2m^2-l\sin^2\theta(m\omega+\omega_sr)^2}{flr^2\sin^2\theta}\phi^2
      -U\left(|\Phi|\right), \nonumber \\
p_\theta  &=p_r-2\frac{f}{lgr^2}
\left[
       r^2\left(\partial_r\phi\right)^2+\left(\partial_\theta\phi\right)^2
\right], 
\nonumber \\
p_\varphi &=\rho-2\frac{f}{lgr^2}
\left[
      r^2\left(\partial_r\phi\right)^2+\left(\partial_\theta\phi\right)^2
\right]
      -2 U\left(|\Phi|\right).
\end{align}
There are two new features induced by rotation worth of notice. 
Firstly, the system becomes what we decide to call \emph{completely anisotropic}, 
meaning $p_r\neq p_\theta\neq p_\varphi$. 
Secondly, the energy density and axial pressure show a cusp due to the absolute value term, 
whose argument changes sign at a point where
\begin{equation}
\label{unsmooth}
-f^2 m^2+l r^2 \sin^2\theta\left(\omega_s+m\frac{\omega}{r}\right)^2=0 \ \ \ 
\Longleftrightarrow \ \ \
m^2 g_{tt}-2m\omega_s g_{t\varphi}+\omega_s^2 g_{\varphi\varphi}=0.
\end{equation}
It is straightforward to understand that this point must occur for all solutions: 
at the origin the only surviving term is $g_{tt}$ (which is negative), while at large distances 
the dominant term is $g_{\varphi\varphi}$ (which is always positive). 
The structure of eq. (\ref{unsmooth}) tempts us to define the parameter vector 
$w^\mu=(m,0,0,-\omega_s)$ that is timelike at the center of the star, 
becomes null at the cusp and finally turns spacelike from that point all the way to infinity. 
We note that the Killing vector field $K' = w^\mu \partial_\mu$ possesses the property
$K' \Phi =0$. At the cusp, this property translates to $j^\mu j_\mu=0$.

The change of sign has also consequences for the eigenvectors. In the region where
$w^\mu$ is timelike, the timelike eigenvector can be expressed as
$\hat{e}^\mu_{\phantom{a}(1-)} = w^\mu/\sqrt{-w^\mu w_\mu}$, whereas in the region where
$w^\mu$ is spacelike, $\hat{e}^\mu_{\phantom{a}(1-)} = v^\mu/\sqrt{-v^\mu v_\mu}$, 
where $v^\mu$ is orthogonal to $w^\mu$, i.~e.~$v^\mu w_\mu =0$.

Let us consider an observer in the comoving frame and identify her four velocity 
with the timelike eigenvector
$U^\mu = \left(U^t,0,0,U^\varphi\right)= \hat{e}^\mu_{\phantom{a}(1-)}$.
The energy $E=-g_{t\mu}U^\mu$ and the angular momentum $L=g_{\varphi\mu} U^\mu$
both will diverge, if she approaches the cusp. Hence we conclude,
that the cusp corresponds to a pathology of the comoving frame. An observer approaching the
cusp needs an infinite amount of energy in order to stay in the comoving frame.
Moreover, the cusp surface envelopes a volume where the four current is spacelike and
 $j_\mu\hat{e}^\mu_{\phantom{a}(1-)}=0$, i.e. the observer measures zero particle number density.

Let us next consider the  charged rotating boson stars.
The equations describing the hydrodynamic quantities are 
lengthy and cumbersome, therefore we do not show them fully here but express them at the 
center of the boson star, $r=0$. Keeping only the leading order contribution for each field, 
we have that at $r\approx 0$ 
\begin{equation}
f\approx f_c,\quad 
l\approx l_c,\quad 
g\approx 1,\quad 
\omega\approx\omega_c r,\quad 
\phi\approx\phi_{c \, m} r^m\sin^m\theta,\quad 
V\approx V_c,\quad 
C\approx C_{c \, 2}r^2\sin^2\theta,
\end{equation}
while at $\theta\approx 0$
\begin{equation}
f\approx f_0(r),\quad 
l\approx l_0(r),\quad 
g\approx 1,\quad 
\omega\approx\omega_0(r),\quad 
\phi\approx\phi_m(r)\theta^m,\quad 
V\approx V_0(r),\quad 
C\approx C_2(r)\theta^2.
\end{equation}

The energy density and pressures at the center then yield
\begin{equation}
\label{asexr}
\rho=2\frac{\phi_{c \, 1}^2 f_c}{l_c}+2\left(\frac{C_{c \, 2} f_c}{l_c}\right)^2, \quad 
p_r=2\left(\frac{C_{c \, 2}f_c}{l_c}\right)^2, \quad 
p_\theta=-\rho, \quad 
p_\varphi=p_r,
\end{equation}
and at $\theta=0$,
\begin{equation}
\rho=2\frac{\phi_1^2(r)f_0(r)}{l_0(r)r^2}
     +\frac{V_0'^2(r)}{2l_0(r)}
     +2\left(\frac{C_2(r)f_0(r)}{l_0(r)r^2}\right)^2, \quad 
p_r=\frac{V_0'^2(r)}{2l_0(r)}
     +2\left(\frac{C_2(r)f_0(r)}{l_0(r)r^2}\right)^2, \quad 
p_\theta=-\rho, \quad 
p_\varphi=p_r.
\end{equation}

The energy density of a rotating boson star with $m=1$ is then nonzero at the origin, 
even when uncharged, although the scalar field vanishes at that point. 
At higher rotational numbers, the charged boson star maintains nonzero density at the 
symmetry axis, as opposed to the uncharged rotating star. We stress the similar behavior 
for the trace of the energy momentum tensor $T^\mu_{\phantom{a}\mu}$, which is always 
negative for $\theta=0$ for $m=1$, but zero on that axis for $m>1$.

\begin{figure}[h!]
\begin{center}
\mbox{\includegraphics[width=0.4\textwidth, angle =0]{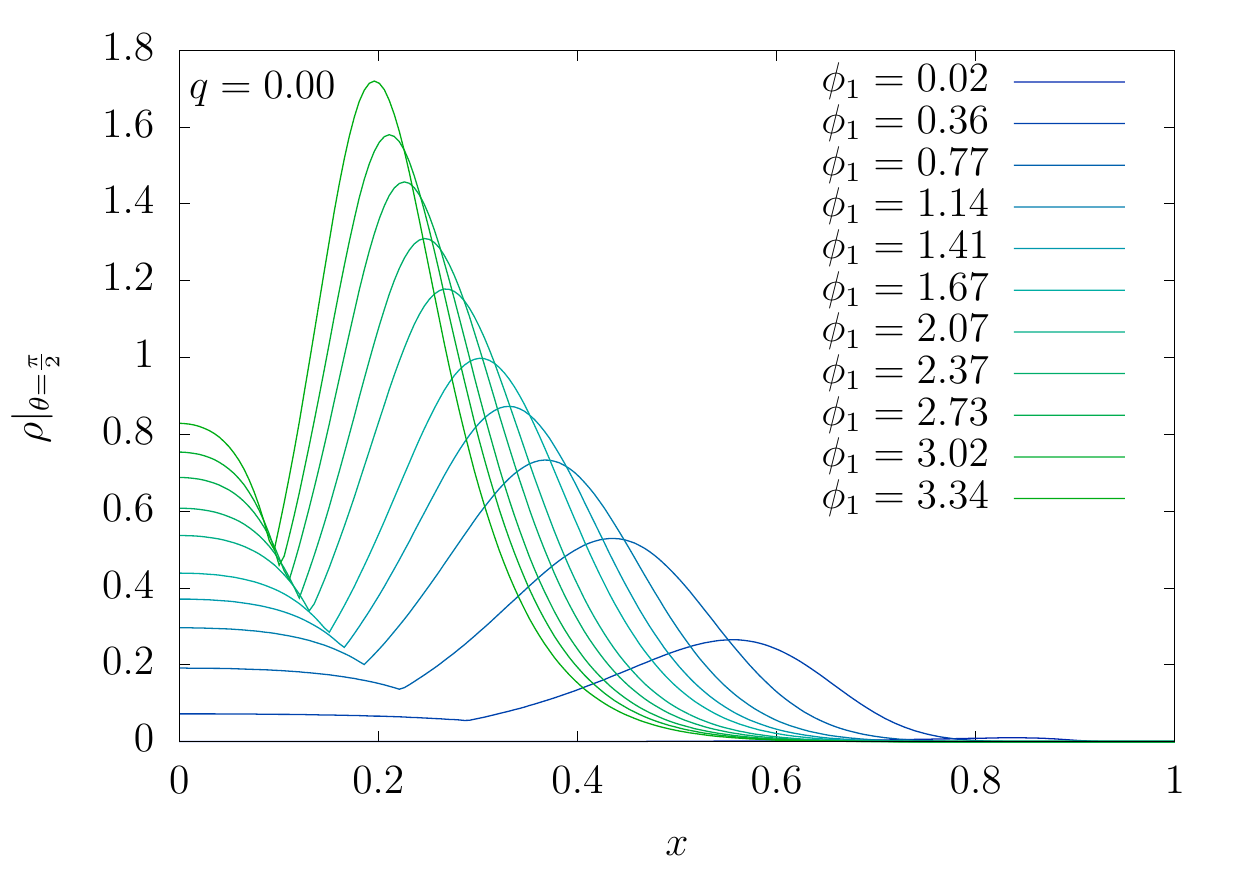}}
\mbox{\includegraphics[width=0.4\textwidth, angle =0]{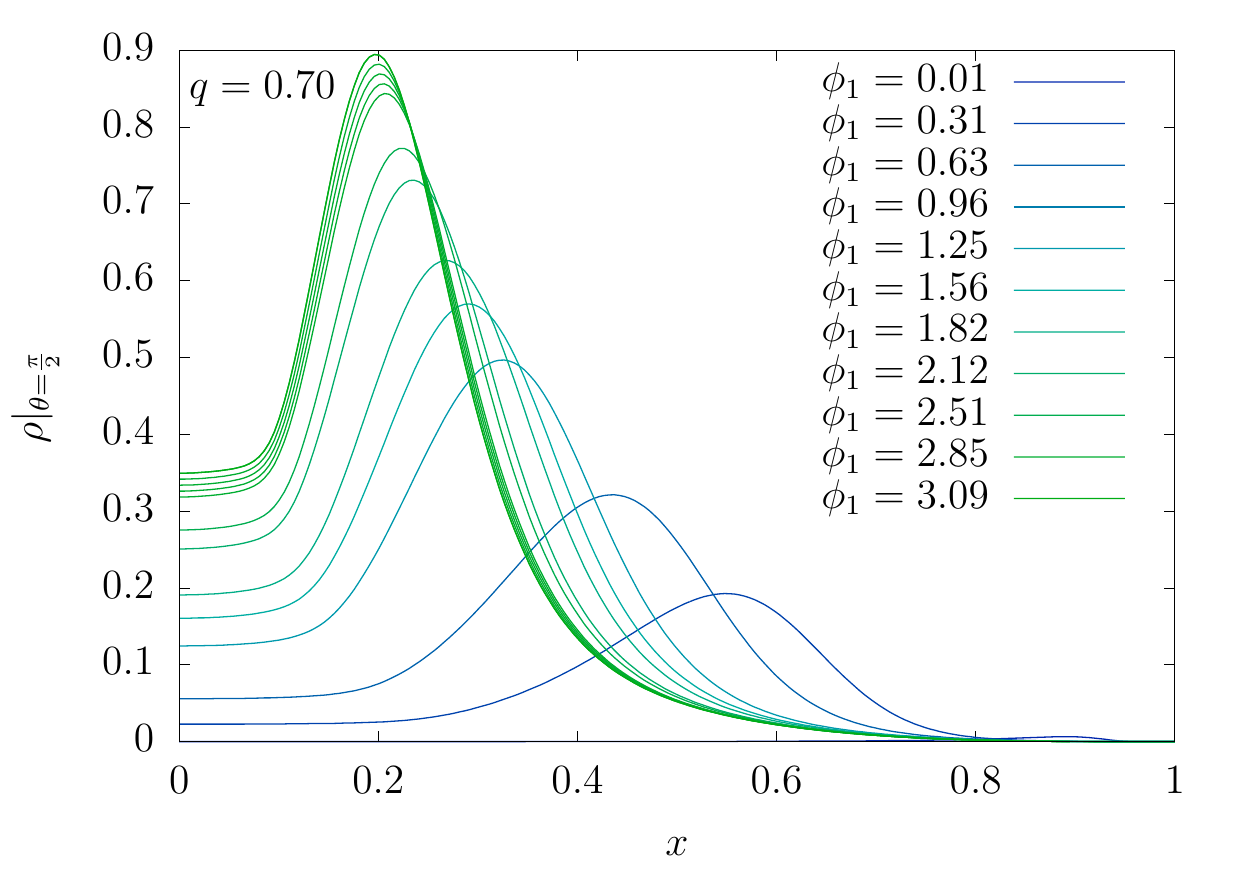}}
\end{center}
\caption{
Profile of the energy density on the equatorial plane for boson stars with rotating 
number $m=1$
\emph{Left:}  Uncharged rotating boson star.  
\emph{Right:} Charged rotating boson star. 
}
\label{peak}
\end{figure}

We show in Fig. \ref{peak} the energy density in the equatorial plane for the uncharged
rotating boson stars (left) and the charged rotating boson stars with gauge coupling parameter
$q=0.7$ (right) for several values of $\phi_1$ and $m=1$. 
In the uncharged case the energy density possesses a local maximum 
at the center, a local minimum corresponding to the cusp, and 
a maximum in the equatorial plane. This is in contrast to the charged case when no cusp is present.
Thus for the charged roating boson stars the energy density possesses a local minimum
at the center and a maximum in the equatorial plane.

With increasing values of the parameter $\phi_1$ 
the magnitudes of the minima and the maxima increase and the locations of the maxima and the cusp
move closer to the center. Note that, in both cases, the central density can be significant high,
despite the vanishing scalar field on the symmetry axis.


%
%

%
\begin{figure}[h!]
\begin{center}
\mbox{\includegraphics[width=0.4\textwidth, angle =0]{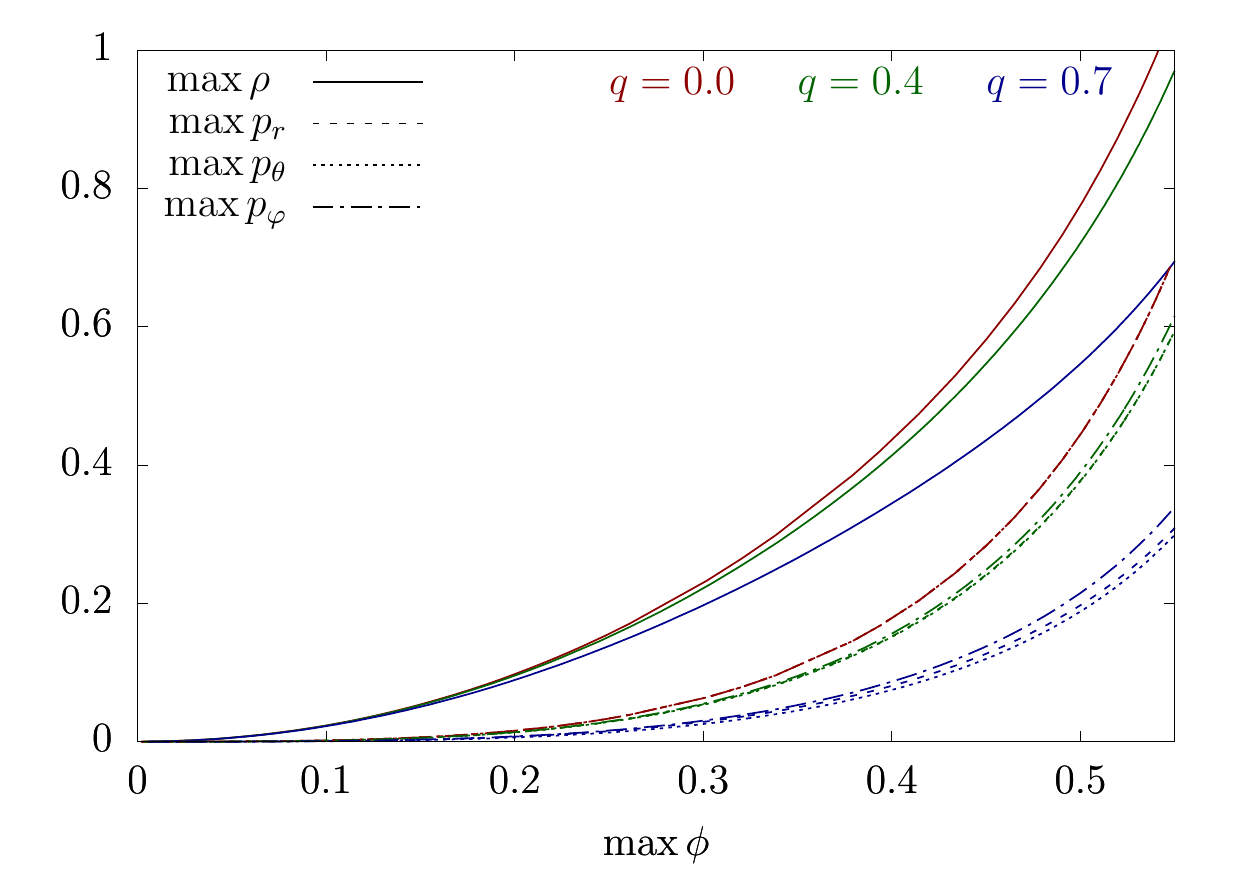}}
\mbox{\includegraphics[width=0.4\textwidth, angle =0]{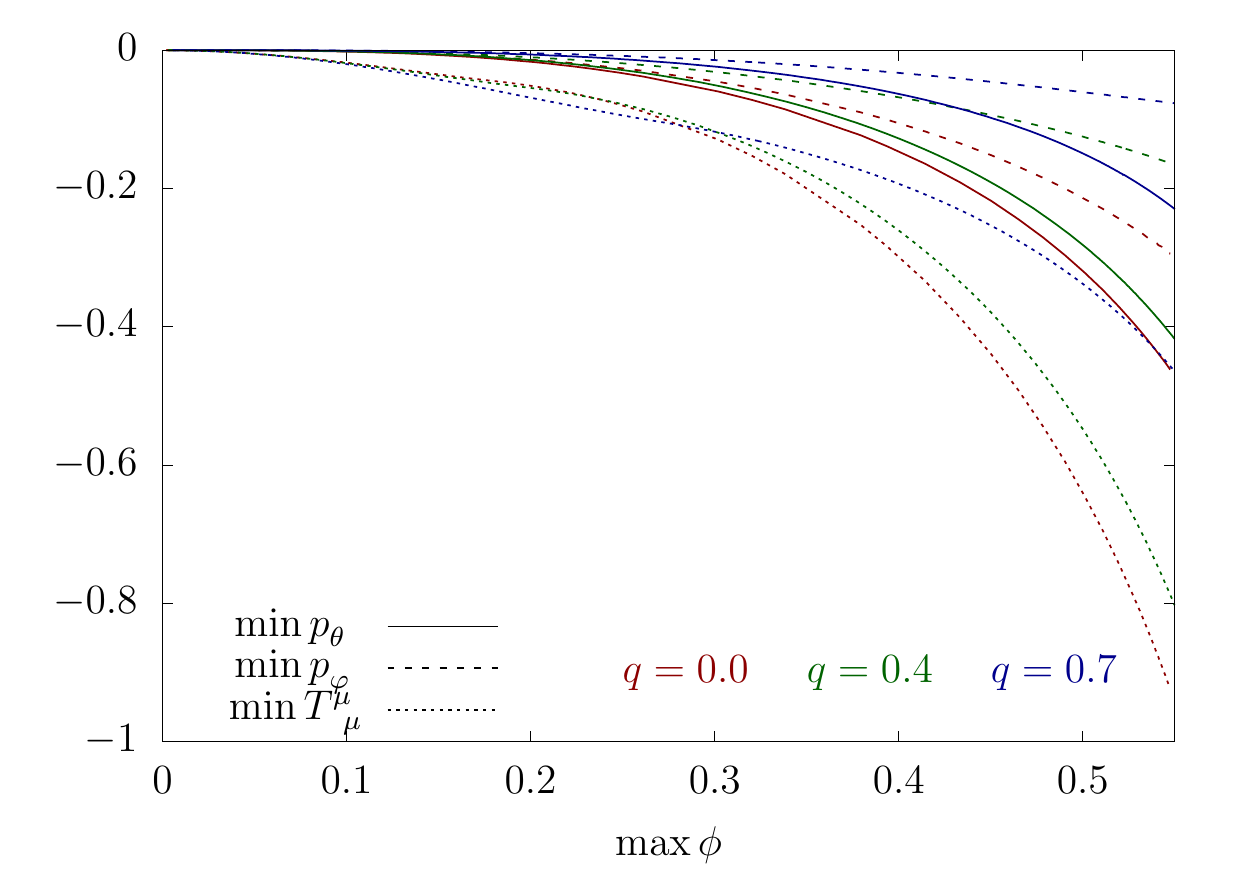}}
\end{center}
\caption{
\emph{Left:} Maximum value of the hydrodynamic quantities for uncharged and 
charges rotating Boson stars as functions of the maximum value of the scalar field 
for each solution. The higher the coupling value, the smaller these quantities become, 
but the lines tend to spread away enhancing the complete anisotropy. 
\emph{Right:} Minimum value of the anisotropic tangential pressures and trace of the energy 
momentum tensor as functions of the maximum value of the field. The minimum value of the energy 
density and radial pressure is zero and therefore not depicted. Again here, 
by increasing the charge, the minimum value is mitigated.
}
\label{pmr}
\end{figure}
%

%
\begin{figure}[h!]
\begin{center}
\mbox{\includegraphics[width=0.4\textwidth, angle =0]{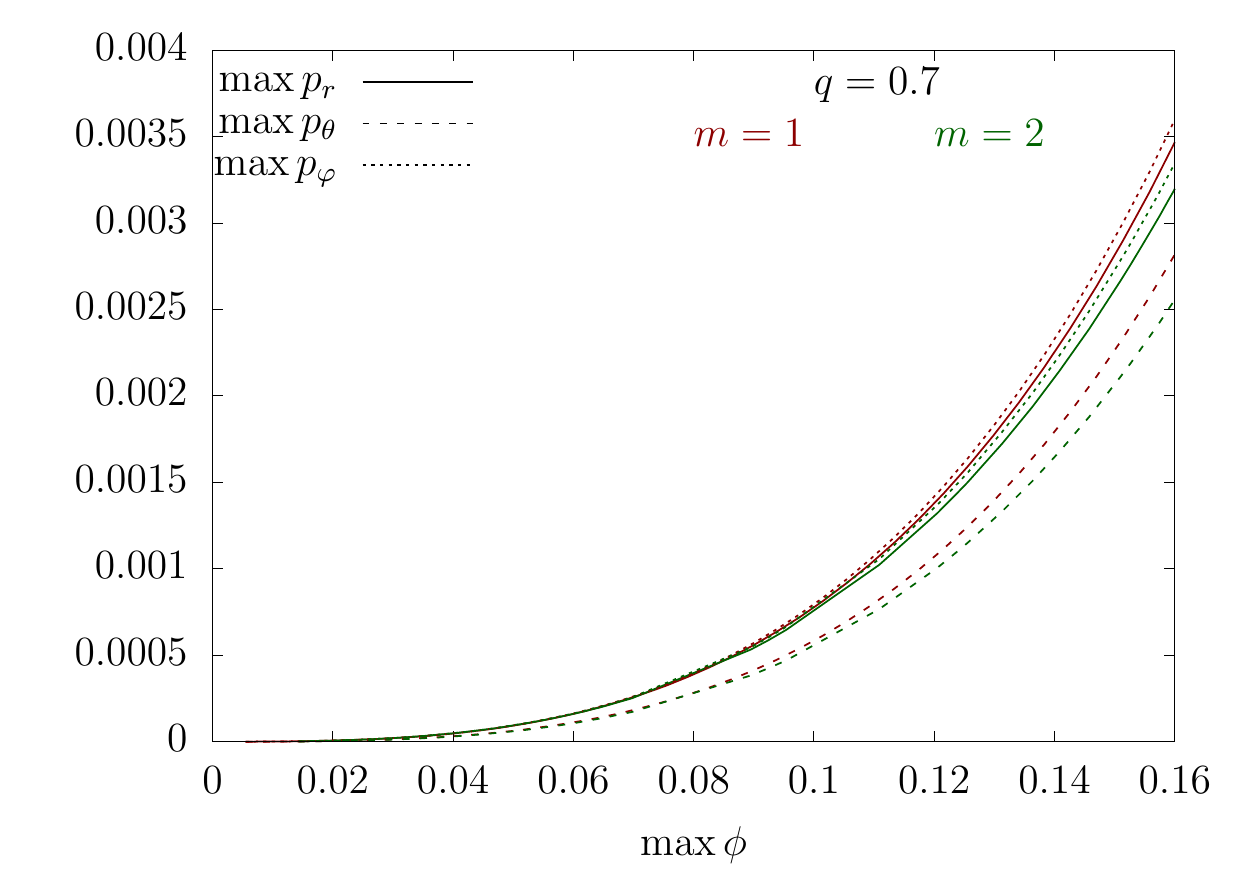}}
\end{center}
\caption{
Maximum of pressures for rotating charged boson stars with rotational numbers $m=1$ and $m=2$ 
as functions of the maximum of the scalar field.Rotation has a similar effect on these 
quantities as does the charge, decreasing their maximum value and slightly pushing these 
curves further apart.
}
\label{rpm1m2}
\end{figure}
In Fig. \ref{pmr} we illustrate how the maximum of the energy density and anisotropic pressures 
vary with respect to the maximum of the scalar field for solutions with different values of the
gauge coupling parameter, 
as well as the minimum of the tangential pressures and trace of the energy momentum tensor. 
The energy density and radial pressure are everywhere positive and their global minimum is therefore zero,
 similarly the trace of the energy momentum tensor has its maximum at zero for 
being everywhere negative defined. The uncharged rotating solution, although already 
completely anisotropic, features for every solution the same maximum value for all three 
pressures. With increasing gauge coupling the maxima and minima of all hydrodynamic quantities 
become smaller, but the curves giving the maximum value of the pressures start to 
separate from each other. For $q=0.4$ we notice that the maximum value of $p_\varphi$ is 
always bigger than those of $p_r$ and $p_\theta$ which lie on the same curve, and 
for $q=0.7$ we notice the $p_\theta$ curve appearing below the maximum radial pressure one. 
We stress that we need to go to very large couplings, near the limiting $q_{crit}$, 
to be able to observe such small deviations.

For completeness, we show in Fig. \ref{rpm1m2} how the maximum values of the pressures 
vary  with the rotational number $m$. The domain of existence of boson stars with $m=2$ is
more restricted as compared  to the case with $m=1$.
The density grows much more rapidly with increasing $\phi_2$, and the maximum of the 
scalar field takes smaller values when compared with $m=1$. As the rotational number 
assumes larger values, the dynamical properties of the boson star tend to be more and 
more dominated by its kinetic terms in the energy momentum tensor. Furthermore, 
increasing $m$ has an analogous effect on the maximum of pressures as does an increasing 
charge, i.e. they decrease in value and the curves become distinct. 
No difference was noted for the 
energy density of minimum of such quantities.

In the three figures that follow, we present three different rotating boson stars for comparison. 
The first boson star is uncharged, $q=0$ and has rotational number $m=1$. 
The second and third both have the same coupling charge, $q=0.7$, 
but the latter possesses rotational number $m=2$. 
Each of these solutions corresponds to the 
one with the maximum mass for fixed $m$ and $q$. 
In these images and the others that will appear later, $z=\bar{r}\cos\theta$ is parallel to the rotation axis and 
$x=\bar{r}\sin\theta$ to the equatorial plane, where $\bar{r}=r/(r+1)$ is the compactified radial coordinate.

\begin{figure}[h!]
\begin{center}
\mbox{
\includegraphics[width=0.4\textwidth, angle =0]{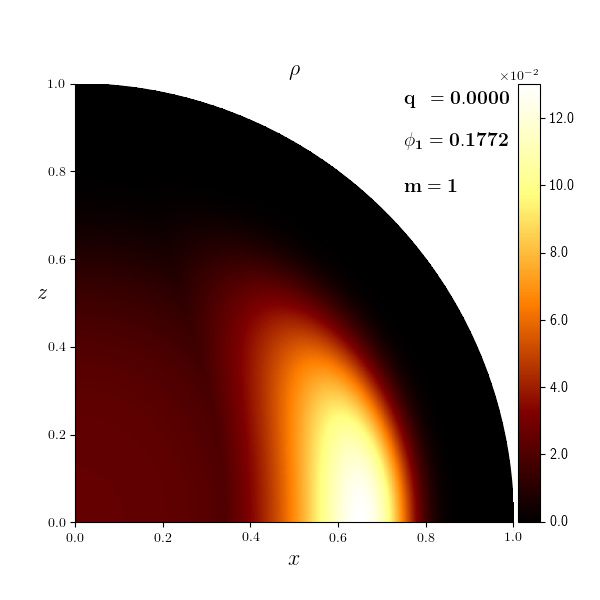}
\includegraphics[width=0.4\textwidth, angle =0]{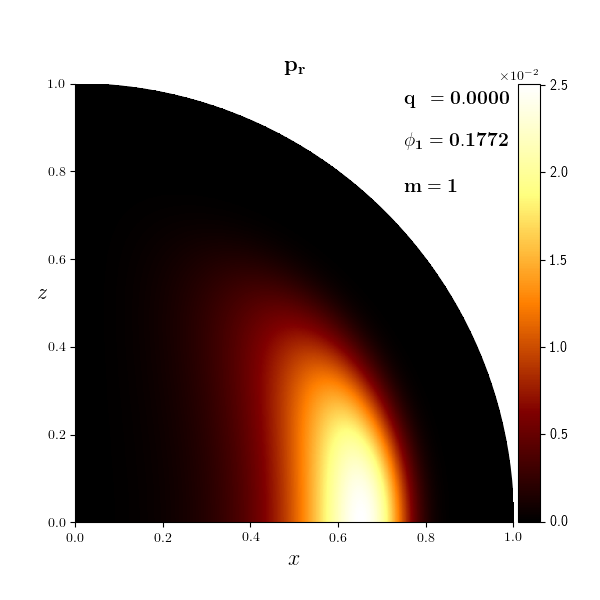}
}\\
\mbox{
\includegraphics[width=0.4\textwidth, angle =0]{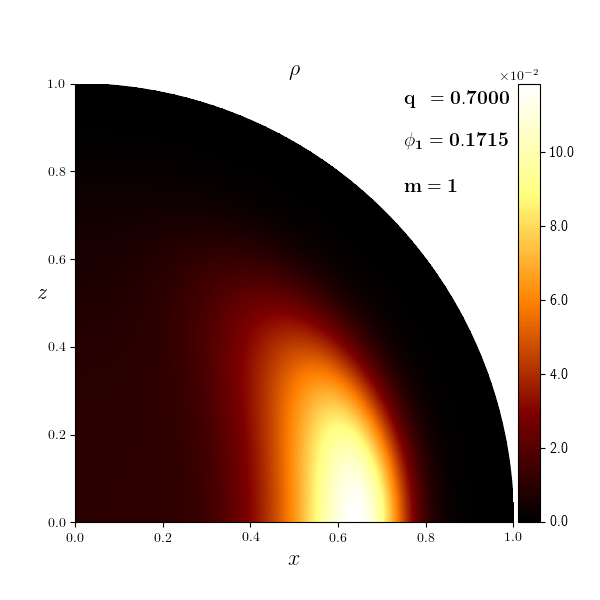}
\includegraphics[width=0.4\textwidth, angle =0]{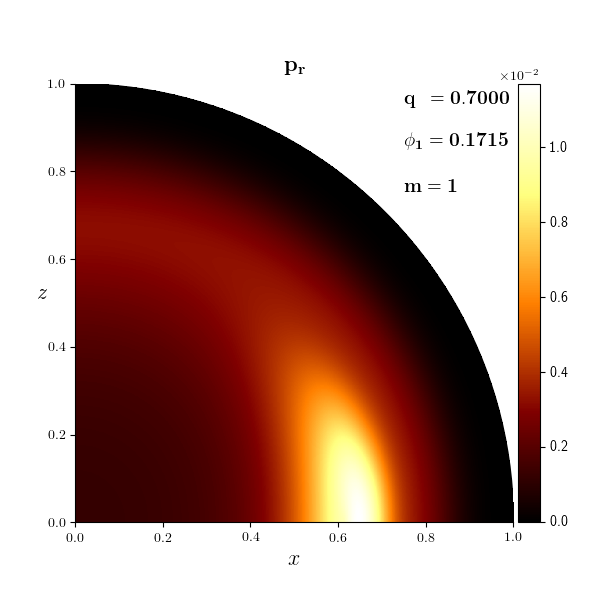}
}\\
\mbox{
\includegraphics[width=0.4\textwidth, angle =0]{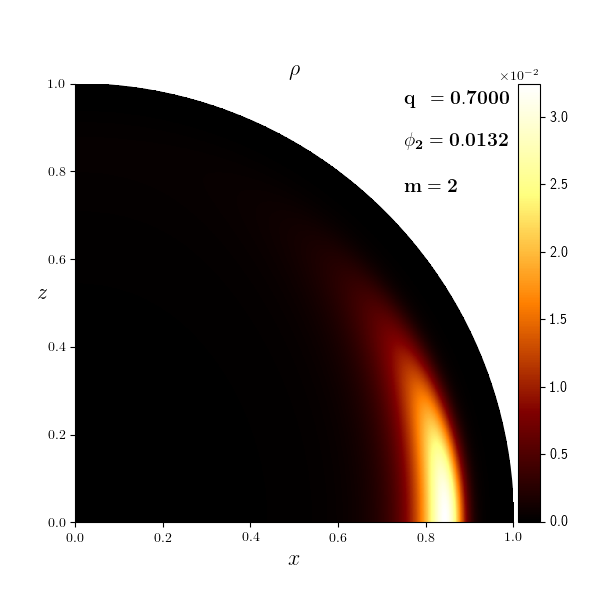}
\includegraphics[width=0.4\textwidth, angle =0]{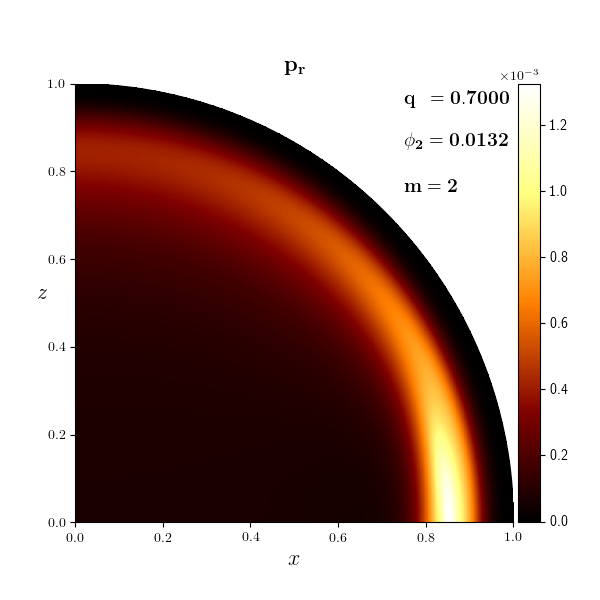}
}
\end{center}
\caption{Energy density and radial pressure as seen by an observer comoving with the fluid 
of a rotating charged boson star for different charges and rotational numbers}
\label{rpr}
\end{figure}

The energy density and radial pressure are depicted in Fig. \ref{rpr}. The comoving observer 
measures the energy density to be a lot higher off center, being negligible at the core for 
the $m=2$ star, as according to eq. (\ref{asexr}) only $\partial^2_rC$ contributes at this point, 
and its value is fairly low. For the uncharged case, we can entertain the kink in the 
energy density discussed above, as its value drops harshly to a local minimum and then 
grows back up to the global maximum with increasing $r$, and we see a thin dark line that 
spans through all the polar coordinate. Furthermore, in this same case the radial pressure 
is zero at the origin and negligible outside the neighborhood of its maximum value. 
The faster rotating star, as seen by the comoving observer, has a thinner density bag 
which sits further away from the origin as one would expect. The profile of the radial 
pressure is quite similar, but we notice how the bag is somewhat spread to smaller values 
of $\theta$.

\begin{figure}[h!]
\begin{center}
\mbox{
\includegraphics[width=0.4\textwidth, angle =0]{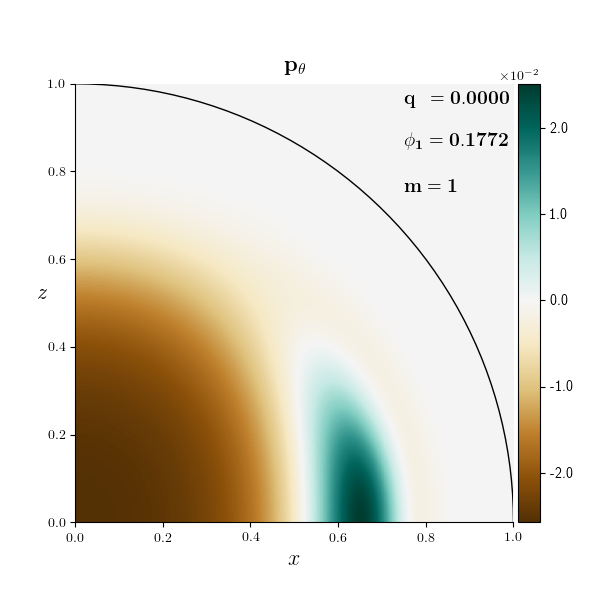}
\includegraphics[width=0.4\textwidth, angle =0]{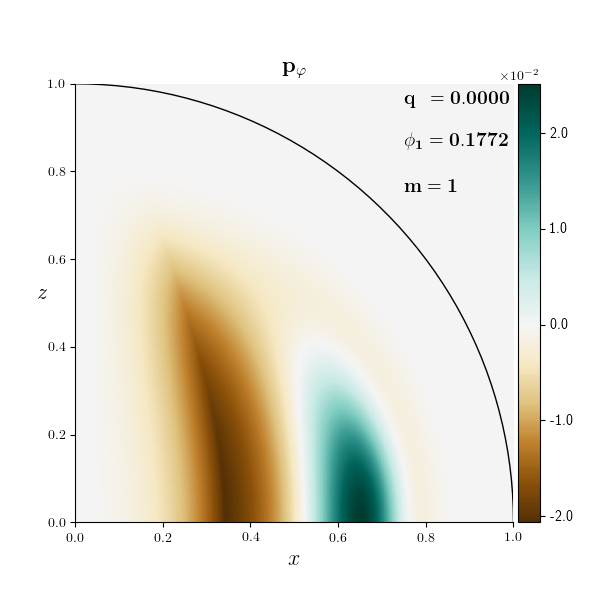}
}\\
\mbox{
\includegraphics[width=0.4\textwidth, angle =0]{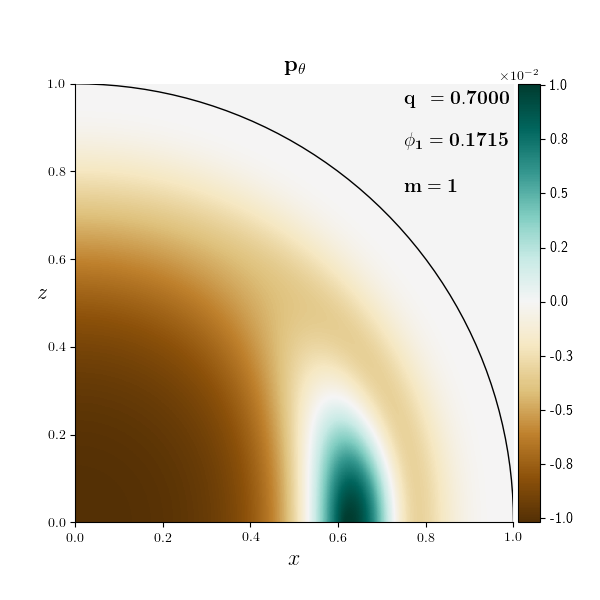}
\includegraphics[width=0.4\textwidth, angle =0]{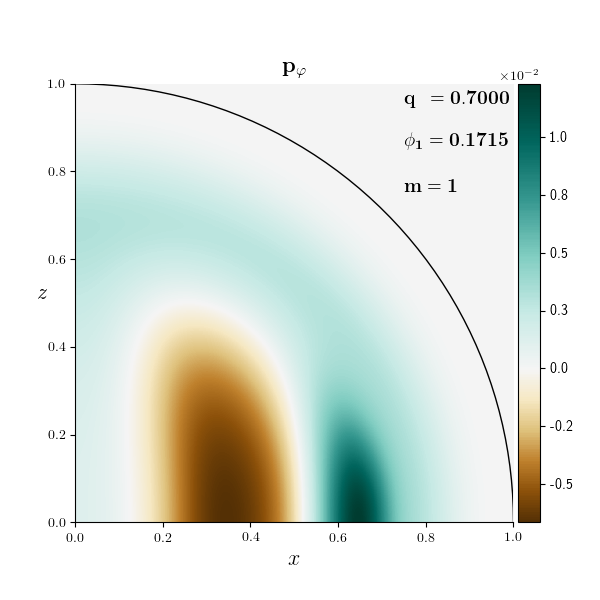}
}\\
\mbox{
\includegraphics[width=0.4\textwidth, angle =0]{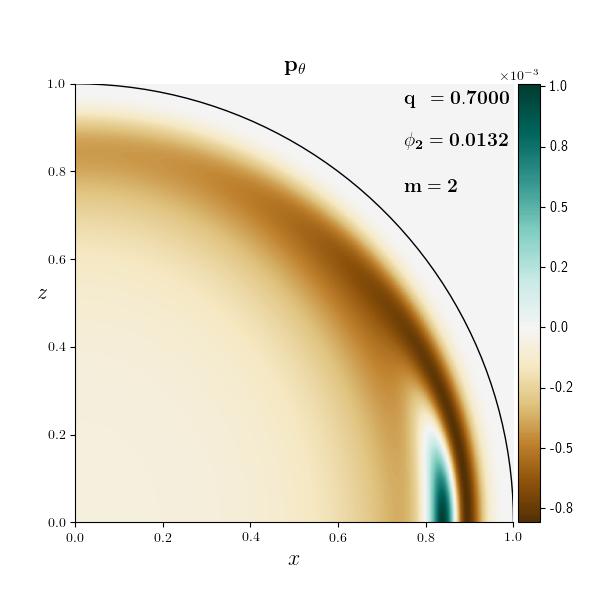}
\includegraphics[width=0.4\textwidth, angle =0]{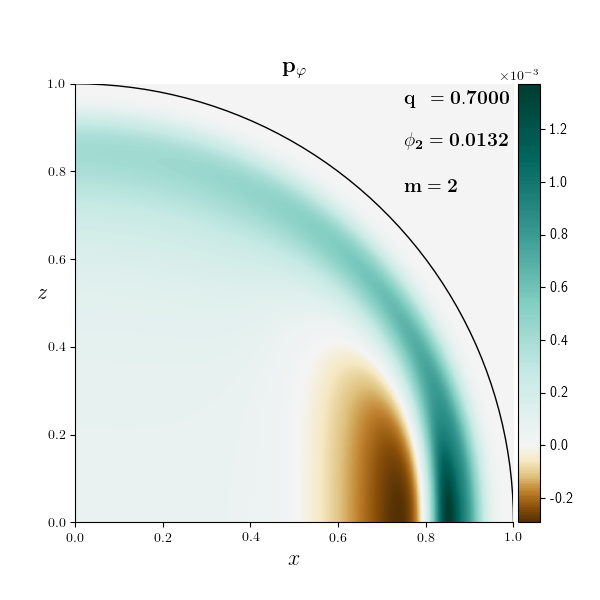}
}
\end{center}
\caption{Polar and axial pressures for the same stars as in \ref{rpr}, 
as measured by the comoving observer.}
\label{tpr}
\end{figure}

In Fig. \ref{tpr}, the two tangential pressures are illustrated, and we are able to 
acknowledge how distinct they are. For $m=1$, the polar pressure is highly negative near 
the center while the axial one is mildly positive for the charged case and zero for $q=0$. 
In all three stars, in the panel for $p_\varphi$, there is an empty, pressureless shell 
which encompasses a region where the sign of the pressure switches. At higher distances 
from the center, the axial pressure becomes once again positive, and we notice that all 
pressures have their maximum value near the point of highest energy density. As before, we 
notice how these quantities distribute themselves over a wider range in the polar coordinate 
for $m=2$, while getting narrower in $r$.

\begin{figure}[h!]
\begin{center}
\mbox{
\includegraphics[width=0.4\textwidth, angle =0]{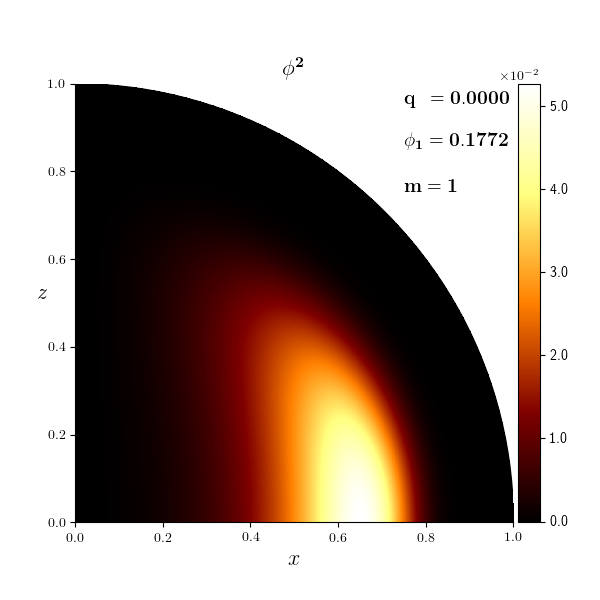}
\includegraphics[width=0.4\textwidth, angle =0]{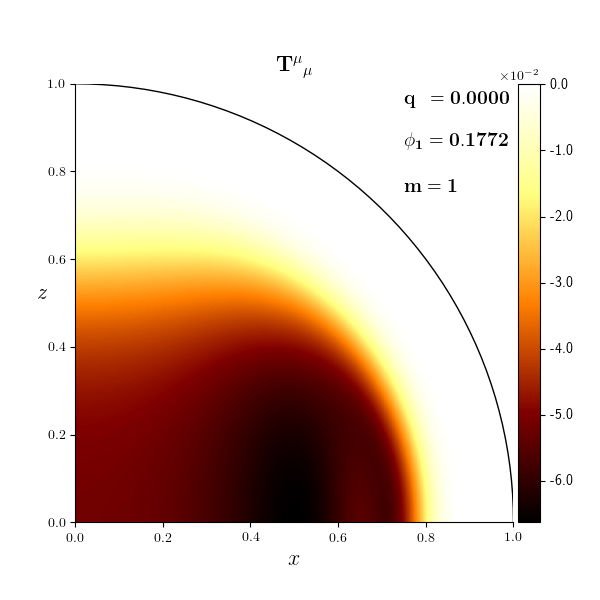}
}\\
\mbox{
\includegraphics[width=0.4\textwidth, angle =0]{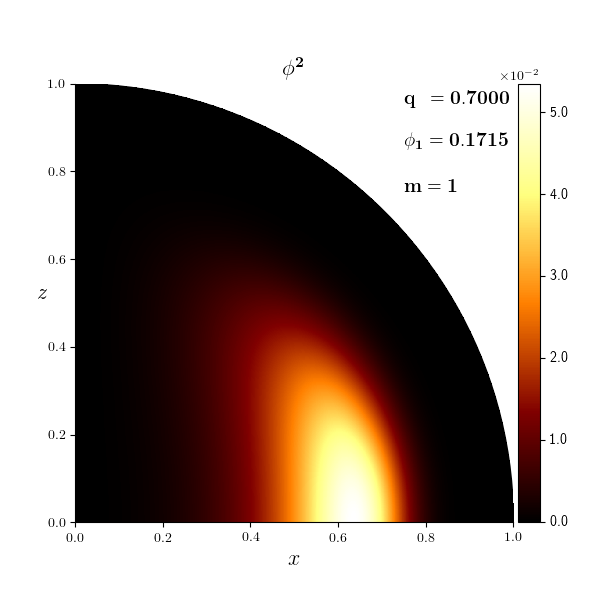}
\includegraphics[width=0.4\textwidth, angle =0]{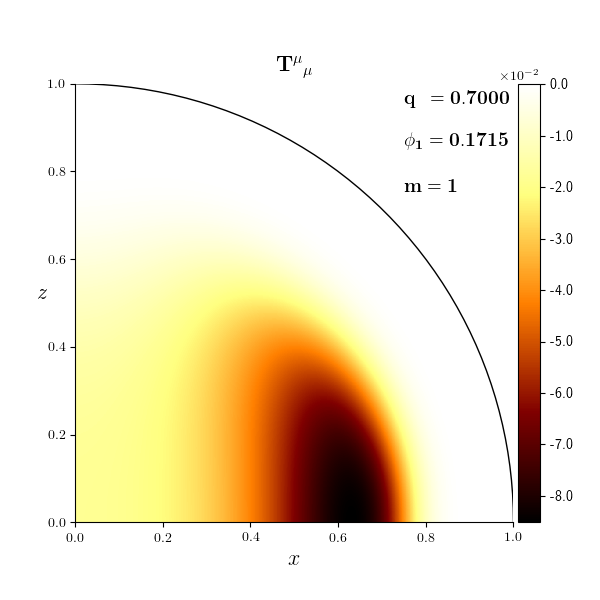}
}\\
\mbox{
\includegraphics[width=0.4\textwidth, angle =0]{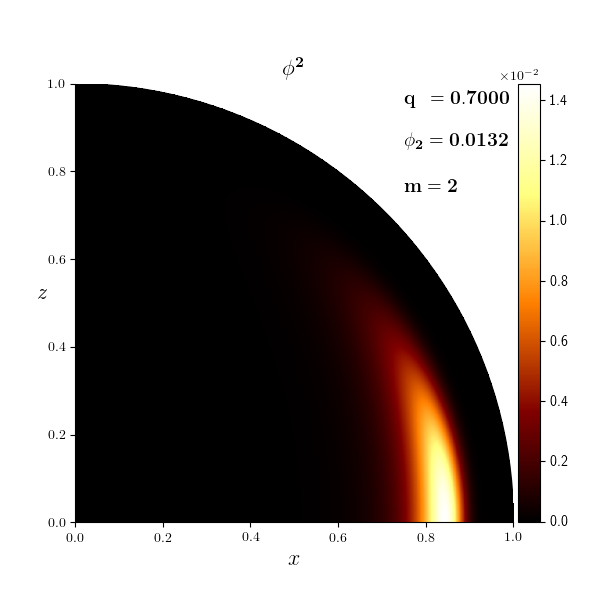}
\includegraphics[width=0.4\textwidth, angle =0]{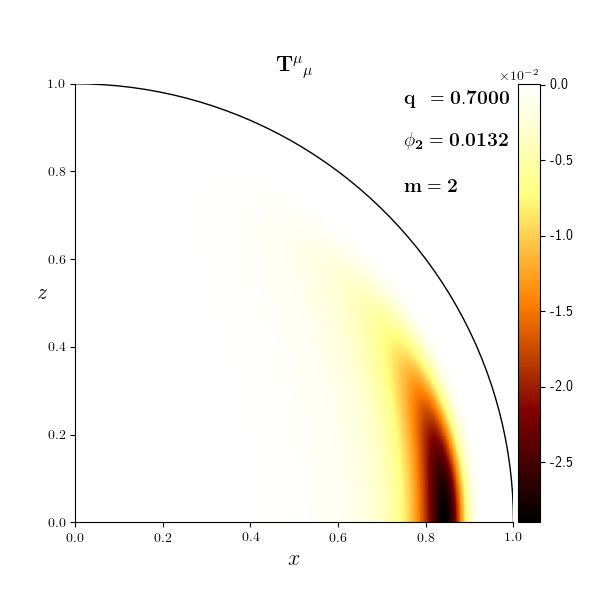}
}
\end{center}
\caption{The squared of the scalar field (photon's mass) and the trace of the energy momentum 
tensor for the same stars as in \ref{rpr} and \ref{tpr}.}
\label{p2tr}
\end{figure}

The scalar field, which sources the electromagnetic field
is draw in Fig. \ref{p2tr} together with the trace of the energy momentum tensor. 
The $\phi^2$ profile has the shape of a torus, as could be anticipated by the boundary 
conditions. The trace of the energy momentum tensor, 
which is zero for the electromagnetic field, takes now negative values.


%
%


%
%

%

\section{Electromagnetic Field}
\label{s4}
\begin{figure}[h!]
\begin{center}
\mbox{
\includegraphics[width=0.4\textwidth, angle =0]{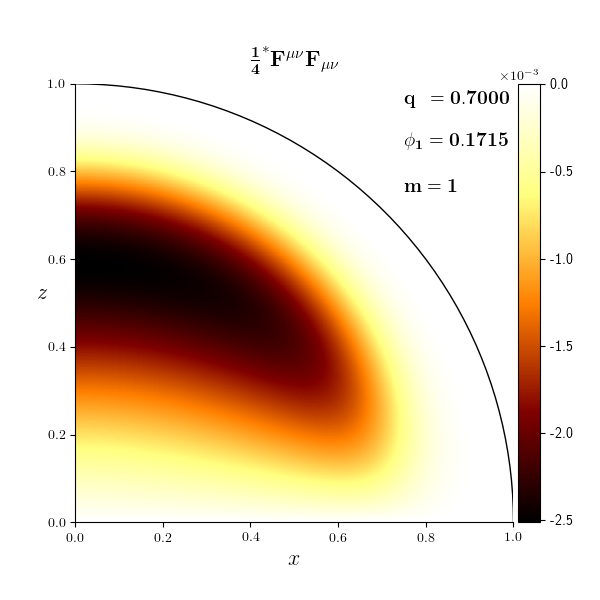}
\includegraphics[width=0.4\textwidth, angle =0]{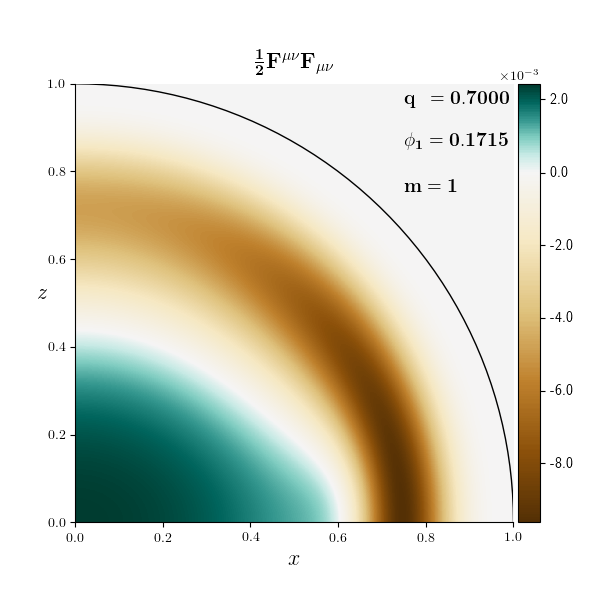}
}\\
\mbox{
\includegraphics[width=0.4\textwidth, angle =0]{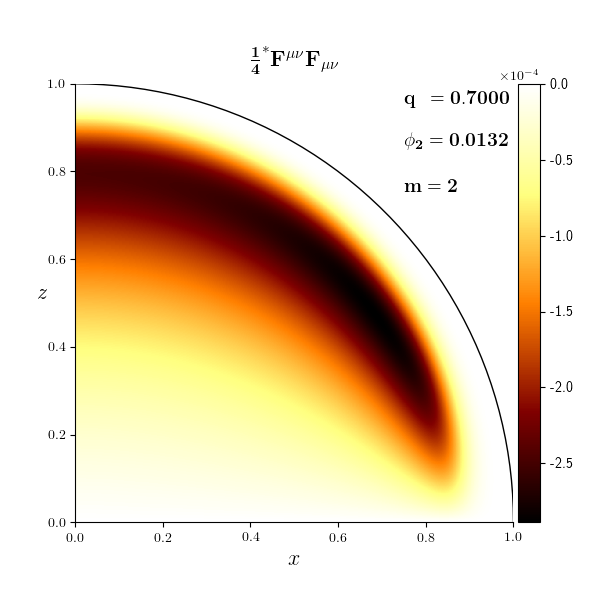}
\includegraphics[width=0.4\textwidth, angle =0]{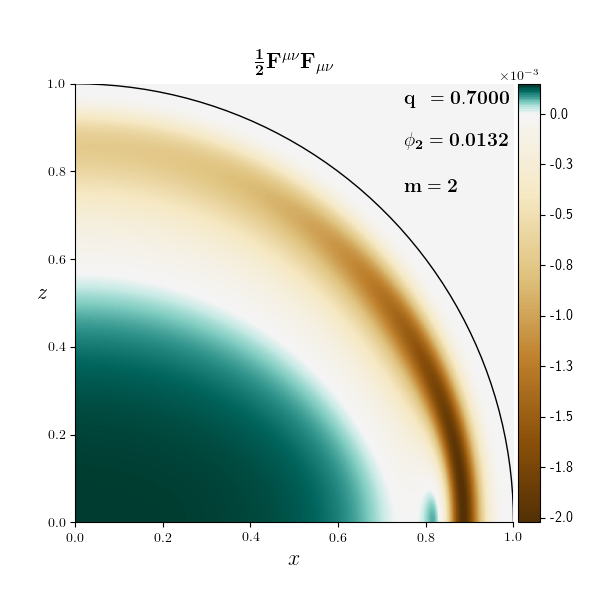}
}\\
\mbox{
\includegraphics[width=0.4\textwidth, angle =0]{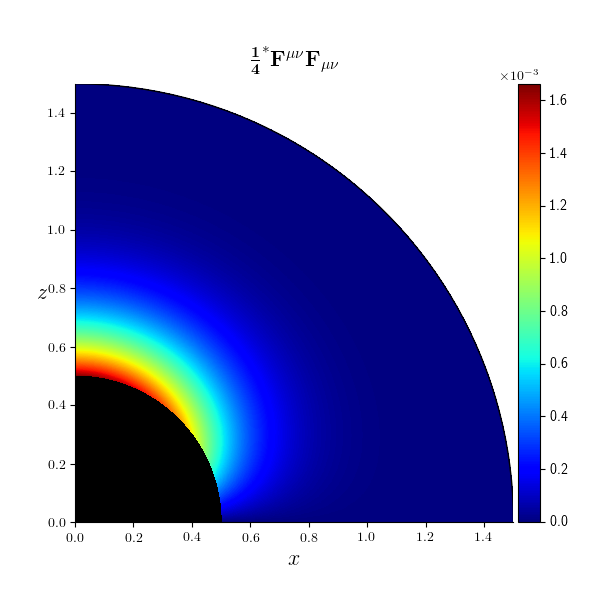}
\includegraphics[width=0.4\textwidth, angle =0]{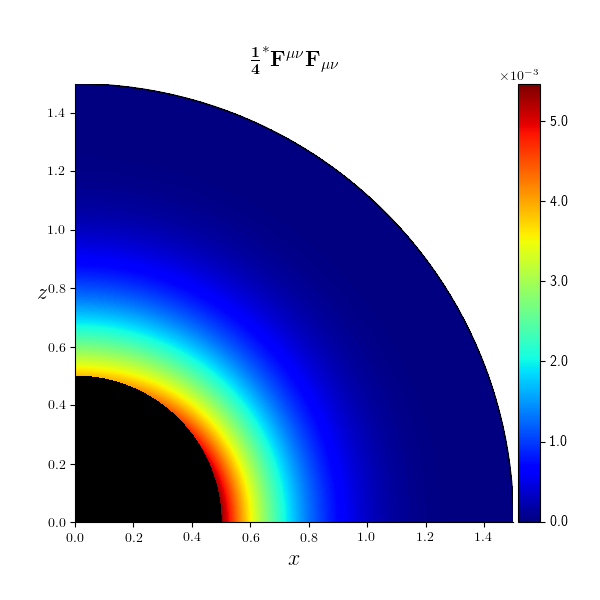}
}
\end{center}
\caption{Invariants of the electromagnetic fields for the two charged boson stars shown in 
the previous figures and a Kerr-Newman black hole (bottom) with the same mass, charge and 
spin as the star on the top panel. For the black hole, only the exterior values are show. 
The black disc covers the inner horizon region.}
\label{invem}
\end{figure}

The two invariants of the electromagnetic field are seen in Fig. \ref{invem}, 
where ${}^{\ast}F_{\mu\nu}=\frac{1}{2}\epsilon_{\mu\nu\sigma\lambda}F^{\sigma\lambda}$ 
is the dual of the field and $\epsilon_{\mu\nu\sigma\lambda}$ is the Levi-Civita tensor. 
The invariants for a Kerr-Newman black hole which possesses the same mass, charge and angular 
momentum as the analyzed boson star with $m=1$, is also given at the bottom, where the black 
disc represents its event horizon and we show only the contracted fields in the exterior region. 
The general behavior is drastically different. Both invariants extend to much larger regions 
in comparison with the Kerr-Newman black hole, since their charge concentrates off center and 
the mains resemblance is the orthogonality between the fields on the equatorial plane, 
which could be anticipated from the boundary conditions we established. Rotating charged 
boson stars feature a region of strong magnetic dominance, which is lacking in rotating 
charged black holes.

%

As a means to visualize the electric and magnetic field individually, 
we need to adopt a reference frame since those are not invariant quantities. 
Therefore, we choose the local inertial frame of a zero angular momentum observer (ZAMO), 
the only one capable of inertially reaching the center of the star, 
see \cite{Grandclement:2014msa,Grould:2017rzz,Collodel:2017end}. The fields are then given by
\begin{equation}
\label{electmag}
E_\mu=F_{\mu\nu}\chi^\nu; \qquad 
B_\mu=-\frac{1}{2}\epsilon_{\mu\nu\sigma\gamma}F^{\nu\sigma}\chi^\gamma,
\end{equation}
where $\chi^\mu$ is the four-velocity of the ZAMO, which reads in general form 
(in a stationary, axisymmetric spacetime),
\begin{equation}
\chi^\mu=\sqrt{-g^{tt}}\left(1,0,0,\frac{g^{t\varphi}}{g^{tt}}\right),
\end{equation}
and one should note that, indeed, $\chi_\varphi=L=0$.

\begin{figure}[h!]
\begin{center}
\mbox{
\includegraphics[width=0.4\textwidth, angle =0]{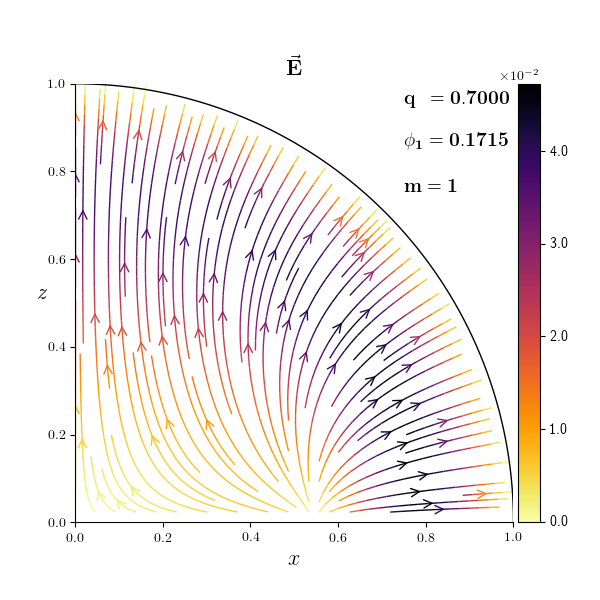}
\includegraphics[width=0.4\textwidth, angle =0]{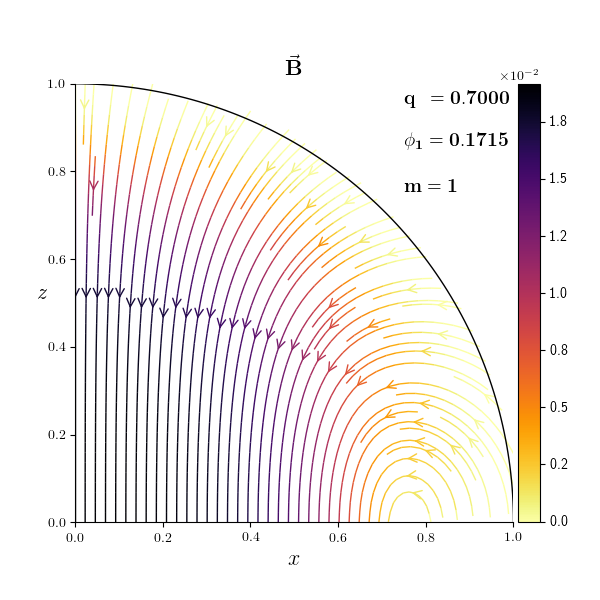}
}\\
\mbox{
\includegraphics[width=0.4\textwidth, angle =0]{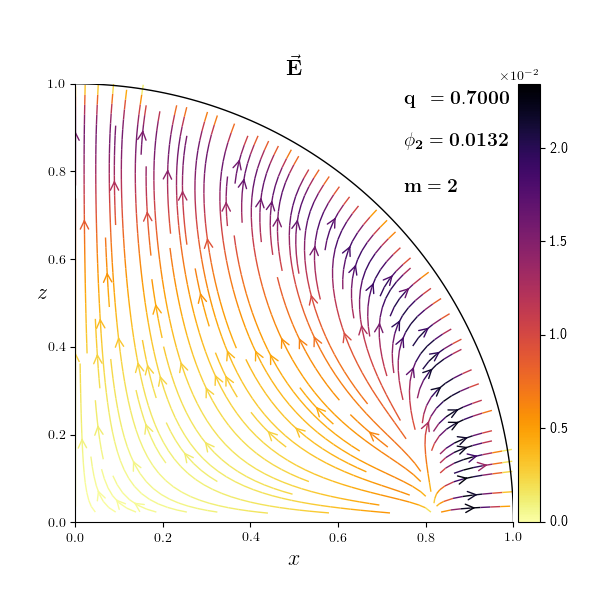}
\includegraphics[width=0.4\textwidth, angle =0]{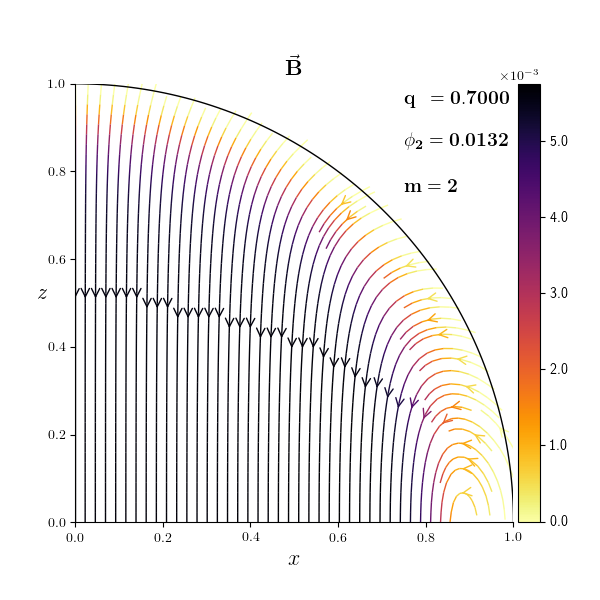}
}
\end{center}
\caption{Electric and magnetic field as measured by the ZAMO.}
\label{ebf}
\end{figure}

These fields, as seen by the ZAMO, are given in Fig. \ref{ebf} for the two previously illustrated 
charged boson stars. The electric field is stronger in a thin shell that encompasses the region 
where the scalar field is maximum, in the \emph{exterior} region. In the \emph{interior} region, 
i.e. for smaller radii than the position of the maximum of the scalar field, the electric 
field is very weak on the equatorial plane. The magnetic field is fairly strong and homogeneous 
in this region. The qualitative behavior of the fields is very similar indeed to that of a 
thick circular loop.


%

\section{Conclusion}
\label{concl}
In the present work we revisited boson stars in their most general form, possessing angular 
momentum and charge. Treating the fields as fluids in a comoving local frame, the hydrodynamic 
quantities were obtained in an unambiguous form. The yielded relationships entertain the 
completely anisotropic character of rotating boson stars, charged or not, that contain three 
different kinds of pressures associated each with a spacelike tetrad basis component. 
We showed that the uncharged rotating boson star has, increasingly with its central density, 
a point where the variation of its energy density diverges, due to an absolute value term in 
its expression, which is also present in its relation for the axial pressure. Furthermore, 
these quantities are not zero at the core, as opposed to the scalar field. As the charge 
coupling increases, the curves describing the maximum and minimum of each of the hydrodynamic variables 
diverge from each other. Unlike most anisotropic stars constructed in an ad-hoc manner, 
the different tangential pressures of a boson star assume at points negative values. 

Measurable entities, such as mass, angular momentum, total charge and magnetic moment were 
also drawn for different configuration sets. As one increases the charge coupling, 
all of these observables take higher values for the same value of the leading order term 
of the scalar field at the origin. Thus, the onset of ergoregions occurs earlier for stars 
with higher charge coupling $q$, terminating the existence of a static ring for 
timelike particles.

The invariants of the electromagnetic field were shown for solutions with different rotation
quantum number $m$, in comparison to a Kerr-Newman black hole with same mass, angular 
momentum and charge as one of the depicted solutions. Even though their order of magnitude 
is the same, the distribution is entirely different thanks to the non trivial topology of 
the scalar field which carries the charge. As seen by a ZAMO, the rotating charged boson 
star produces a fairly homogeneous magnetic field in a neighborhood of the equatorial plane 
in a region between the center of the star and the densest part of the torus.

\section{Acknowledgments}

We would like to acknowledge support by the DFG Research Training Group 1620
{\sl Models of Gravity} as well as by FP7, Marie Curie Actions, People,
International Research Staff Exchange Scheme (IRSES-606096), 
COST Action CA16104 {\sl GWverse}.


\begin{thebibliography}{unsrtnat}



\bibitem{Kaup:1968zz} 
  D.~J.~Kaup,
  Phys.\ Rev.\  {\bf 172}, 1331 (1968).
  doi:10.1103/PhysRev.172.1331

\bibitem{Feinblum:1968nwc} 
  D.~A.~Feinblum and W.~A.~McKinley,
  Phys.\ Rev.\  {\bf 168}, no. 5, 1445 (1968).
  doi:10.1103/PhysRev.168.1445


\bibitem{Ruffini:1969qy} 
  R.~Ruffini and S.~Bonazzola,
  Phys.\ Rev.\  {\bf 187}, 1767 (1969).
  doi:10.1103/PhysRev.187.1767

\bibitem{Jetzer:1989av} 
  P.~Jetzer and J.~J.~van der Bij,
  Phys.\ Lett.\ B {\bf 227}, 341 (1989).
  doi:10.1016/0370-2693(89)90941-6

\bibitem{Yoshida:1997qf} 
  S.~Yoshida and Y.~Eriguchi,
  Phys.\ Rev.\ D {\bf 56}, 762 (1997).
  doi:10.1103/PhysRevD.56.762

\bibitem{Mielke:1997ag} 
  E.~W.~Mielke and F.~E.~Schunck,
  In *Pronin, P. (ed.): Sardanashvili, G. (ed.): Gravity, particles, and space-time* 391-420



\bibitem{Kobayashi:1994qi} 
  Y.~Kobayashi, M.~Kasai and T.~Futamase,
  Phys.\ Rev.\ D {\bf 50}, 7721 (1994).
  doi:10.1103/PhysRevD.50.7721

\bibitem{Colpi:1986ye} 
  M.~Colpi, S.~L.~Shapiro and I.~Wasserman,
  Phys.\ Rev.\ Lett.\  {\bf 57}, 2485 (1986).
  doi:10.1103/PhysRevLett.57.2485

\bibitem{Friedberg:1986tq} 
  R.~Friedberg, T.~D.~Lee and Y.~Pang,
  Phys.\ Rev.\ D {\bf 35}, 3658 (1987).
  doi:10.1103/PhysRevD.35.3658

\bibitem{Deppert:1979au} 
  W.~Deppert and E.~W.~Mielke,
  Phys.\ Rev.\ D {\bf 20}, 1303 (1979).
  doi:10.1103/PhysRevD.20.1303

\bibitem{Coleman:1985ki} 
  S.~R.~Coleman,
  Nucl.\ Phys.\ B {\bf 262}, 263 (1985)
  Erratum: [Nucl.\ Phys.\ B {\bf 269}, 744 (1986)].
  doi:10.1016/0550-3213(85)90286-X, 10.1016/0550-3213(86)90520-1

\bibitem{Barranco:2010ib} 
  J.~Barranco and A.~Bernal,
  Phys.\ Rev.\ D {\bf 83}, 043525 (2011)
  doi:10.1103/PhysRevD.83.043525
  [arXiv:1001.1769 [astro-ph.CO]].


\bibitem{Mielke:1980sa} 
  E.~W.~Mielke and R.~Scherzer,
  Phys.\ Rev.\ D {\bf 24}, 2111 (1981).
  doi:10.1103/PhysRevD.24.2111

\bibitem{Ryan:1996nk} 
  F.~D.~Ryan,
  Phys.\ Rev.\ D {\bf 55}, 6081 (1997).
  doi:10.1103/PhysRevD.55.6081
\bibitem{Schunck:1996he} 
  F.~E.~Schunck and E.~W.~Mielke,
  Phys.\ Lett.\ A {\bf 249}, 389 (1998).
  doi:10.1016/S0375-9601(98)00778-6
\bibitem{Schunck:1999pm} 
  F.~E.~Schunck and E.~W.~Mielke,
  Gen.\ Rel.\ Grav.\  {\bf 31}, 787 (1999).
  doi:10.1023/A:1026673918588
\bibitem{Mielke:2000mh}
  E.~W.~Mielke and F.~E.~Schunck,
  Nucl.\ Phys.\ B {\bf 564}, 185 (2000)
  doi:10.1016/S0550-3213(99)00492-7
  [gr-qc/0001061].
\bibitem{Kleihaus:2005me}  
  B.~Kleihaus, J.~Kunz and M.~List,
  Phys.\ Rev.\ D {\bf 72}, 064002 (2005)
  doi:10.1103/PhysRevD.72.064002
  [gr-qc/0505143].
\bibitem{Kleihaus:2007vk} 
  B.~Kleihaus, J.~Kunz, M.~List and I.~Schaffer,
  Phys.\ Rev.\ D {\bf 77}, 064025 (2008)
  doi:10.1103/PhysRevD.77.064025
  [arXiv:0712.3742 [gr-qc]].
\bibitem{Hartmann:2010pm} 
  B.~Hartmann, B.~Kleihaus, J.~Kunz and M.~List,
  Phys.\ Rev.\ D {\bf 82}, 084022 (2010)
  doi:10.1103/PhysRevD.82.084022
  [arXiv:1008.3137 [gr-qc]].
\bibitem{Kleihaus:2011sx} 
  B.~Kleihaus, J.~Kunz and S.~Schneider,
  Phys.\ Rev.\ D {\bf 85}, 024045 (2012)
  doi:10.1103/PhysRevD.85.024045
  [arXiv:1109.5858 [gr-qc]].
\bibitem{Collodel:2017biu}  
  L.~G.~Collodel, B.~Kleihaus and J.~Kunz,
  Phys.\ Rev.\ D {\bf 96}, no. 8, 084066 (2017)
  doi:10.1103/PhysRevD.96.084066
  [arXiv:1708.02057 [gr-qc]].

\bibitem{Grandclement:2014msa}
  P.~Grandcl\'{e}ment, C.~Somé and E.~Gourgoulhon,
  Phys.\ Rev.\ D {\bf 90}, 024068 (2014)
  doi:10.1103/PhysRevD.90.024068
\bibitem{Grould:2017rzz} 
  M.~Grould, Z.~Meliani, F.~H.~Vincent, P.~Grandcl\'{e}ment and E.~Gourgoulhon,
  Class.\ Quant.\ Grav.\  {\bf 34}, 215007 (2017) 
  doi:10.1088/1361-6382/aa8d39

\bibitem{Collodel:2017end} 
  L.~G.~Collodel, B.~Kleihaus and J.~Kunz,
  Phys.\ Rev.\ Lett.\  {\bf 120}, no. 20, 201103 (2018)
  doi:10.1103/PhysRevLett.120.201103
  [arXiv:1711.05191 [gr-qc]].


\bibitem{Jetzer:1989us} 
  P.~Jetzer,
  Phys.\ Lett.\ B {\bf 231}, 433 (1989).
  doi:10.1016/0370-2693(89)90689-8
\bibitem{Jetzer:1990wr} 
  P.~Jetzer,
  CERN-TH-5681/90.
\bibitem{Jetzer:1993nk} 
  P.~Jetzer, P.~Liljenberg and B.~S.~Skagerstam,
  Astropart.\ Phys.\  {\bf 1}, 429 (1993)
  doi:10.1016/0927-6505(93)90008-2
  [astro-ph/9305014].
\bibitem{Kleihaus:2009kr} 
  B.~Kleihaus, J.~Kunz, C.~Lammerzahl and M.~List,
  Phys.\ Lett.\ B {\bf 675}, 102 (2009)
  doi:10.1016/j.physletb.2009.03.066
  [arXiv:0902.4799 [gr-qc]].
\bibitem{Kleihaus:2010ep} 
  B.~Kleihaus, J.~Kunz, C.~Lammerzahl and M.~List,
  Phys.\ Rev.\ D {\bf 82}, 104050 (2010)
  doi:10.1103/PhysRevD.82.104050
  [arXiv:1007.1630 [gr-qc]].
\bibitem{Pugliese:2013gsa} 
  D.~Pugliese, H.~Quevedo, J.~A.~Rueda H. and R.~Ruffini,
  Phys.\ Rev.\ D {\bf 88}, 024053 (2013)
  doi:10.1103/PhysRevD.88.024053
  [arXiv:1305.4241 [astro-ph.HE]].
\bibitem{Kan:2017rqk} 
  N.~Kan and K.~Shiraishi,
  Eur.\ Phys.\ J.\ C {\bf 78}, no. 3, 257 (2018)
  doi:10.1140/epjc/s10052-018-5745-9
  [arXiv:1709.00157 [gr-qc]].

\bibitem{Brihaye:2009dx} 
  Y.~Brihaye, T.~Caebergs and T.~Delsate,
  arXiv:0907.0913 [gr-qc].

\bibitem{Kichakova:2013sza} 
  O.~Kichakova, J.~Kunz and E.~Radu,
  Phys.\ Lett.\ B {\bf 728}, 328 (2014)
  doi:10.1016/j.physletb.2013.11.061
  [arXiv:1310.5434 [gr-qc]].







\bibitem{Volkov:2002aj} 
  M.~S.~Volkov and E.~Wohnert,
  Phys.\ Rev.\ D {\bf 66}, 085003 (2002)
  doi:10.1103/PhysRevD.66.085003
  [hep-th/0205157].






\bibitem{UrenaLopez:2001tw} 
  L.~A.~Urena-Lopez,
  Class.\ Quant.\ Grav.\  {\bf 19}, 2617 (2002)
  doi:10.1088/0264-9381/19/10/307
  [gr-qc/0104093].

\bibitem{Delgado:2016jxq} 
  J.~F.~M.~Delgado, C.~A.~R.~Herdeiro, E.~Radu and H.~Runarsson,
  Phys.\ Lett.\ B {\bf 761}, 234 (2016)
  doi:10.1016/j.physletb.2016.08.032
  [arXiv:1608.00631 [gr-qc]].

\bibitem{Radu:2008pp} 
  E.~Radu and M.~S.~Volkov,
  Phys.\ Rept.\  {\bf 468}, 101 (2008)
  doi:10.1016/j.physrep.2008.07.002
  [arXiv:0804.1357 [hep-th]].

\end{thebibliography}
\end{document}